\documentclass[letterpaper]{JHEP3}
\usepackage{epsf,epsfig}
\usepackage{subfigure}
\usepackage{graphicx}
\usepackage{amsmath}
\usepackage{bbold}
\usepackage{axodraw}
\usepackage{slashed}

\newcommand\rep\mathbf
\DeclareMathOperator\Tr{Tr}
\newcommand\trans{\mathrm T}
\newcommand\MSbar{\ensuremath{\overline{\text{MS}}}}
\newcommand\fold\otimes

\newcommand\dashlength5 

\preprint{MADPH--09-1536}

\title{\vspace*{.75in}
QCD Corrections to Scalar Diquark Production \\
at Hadron Colliders}

\author{Tao Han,  Ian Lewis, and 
Thomas McElmurry\footnote{Address after Oct. 2009: Department of Physics, Brookhaven National Laboratory, Upton, NY 11973, USA}~\\
 Department of Physics, University of Wisconsin, Madison, WI 53706, U.S.A.\\
 Email:  {\rm than@hep.wisc.edu,\ ilewis@wisc.edu,\ mcelmurry@hep.wisc.edu }}

\abstract{
We calculate the next-to-leading order QCD corrections to quark-quark annihilation to a scalar 
resonant state (``diquark'') in a color representation of antitriplet or sextet at the Tevatron and LHC 
energies. 
At the LHC, we find the enhancement ($K$-factor) for the antitriplet diquark is typically about 1.31--1.35,
 and for the sextet diquark is about 1.22--1.32 for initial-state valence quarks.
The full transverse-momentum spectrum for the diquarks is also calculated at the LHC
by performing the soft gluon resummation to the leading logarithm and all orders in the
strong coupling. }

\begin{document}
\section{Introduction}

Hadron colliders, such as the Fermilab Tevatron and the CERN Large Hadron Collider (LHC), 
bring us to the high-energy frontier for new discoveries of physics beyond the Standard Model (SM).
Yet they are QCD machines: 
The strong interaction of the colored partons, quarks and gluons, leads to overwhelmingly dominant 
event rates in hadronic collisions. 
Thus any new particles participating in the QCD interaction will be strongly produced with favorable rates.

It is often said that the LHC is a ``gluon factory,'' as initial states involving gluons are the major contributors to 
the majority of the events, particularly for relatively light final states.
On the other hand, due to the abundance of valence quarks in the proton at large momentum fraction $x$, production via initial-state quarks becomes more important for rather heavy particles. 
Since the LHC is a $pp$ collider, production of heavy particles in $qq$ collisions will be significantly enhanced by the valence-valence component of the initial state.
A simple group theory consideration indicates that quark-quark scattering can produce bosonic color states 
$\rep6\oplus\rep{\bar3}$. Particles
in these color representations would be rather exotic states beyond the SM spectrum.

Incidentally, there are indeed theoretical motivations to consider those exotic color states.
Examples include scalar quarks in supersymmetric (SUSY) theories with $R$-parity violation \cite{Barbier:2004ez}, and color-sextet scalar diquarks \cite{Mohapatra:2007af}.%
\footnote{In this work, the word ``diquark'' refers not to a bound state of two quarks, but rather to a fundamental boson carrying the quantum numbers of two quarks.}
Diquark states of exotic quantum numbers have recently been classified in Ref.~\cite{DelNobile:2009st}.  Resonant production of color-antitriplet scalars and vectors at the LHC have been studied in Refs.~\cite{Atag:1998xq,Arik:2001bc,Cakir:2005iw}.
Pair production of color-sextet scalars at the LHC has been studied in Ref.~\cite{Chen:2008hh}.
In this paper we consider the single production of these colored scalar states at hadron colliders including the next-to-leading order (NLO) QCD corrections.
We also present their transverse momentum spectrum by including the leading-log soft gluon resummation to all orders in the strong coupling.  These calculations have been done previously for single scalar top production in $R$-parity violating SUSY \cite{Plehn:2000be}; we have expanded upon that work.

\section{Model And Constraints}
In this paper we are not concerned about the electroweak structure of the diquark, but for completeness Table \ref{qnum.TAB} lists the possible electroweak quantum numbers of the diquark and its coupling to standard model quarks.  Q is the $SU(2)_L$ quark doublet, and U(D) is the up(down)-type $SU(2)_L$ quark singlet.  
The diquark may be either an antitriplet or a sextet under $SU(3)_C$, independently of the electroweak quantum numbers.  Also, the diquark may be either a scalar or vector boson.  For simplicity and due to theoretical motivations \cite{Barbier:2004ez,Mohapatra:2007af}, we concentrate on the scalar case. 
\begin{table}[htb]
\caption{Electroweak quantum numbers of possible diquark states and their allowed couplings to SM quarks.  In each case the $SU(3)_C$ quantum number may be either $\rep{\bar3}$ or $\rep6$.
Under $SU(3)_C\times SU(2)_L\times U(1)_Y$, Q has the quantum numbers $(\rep3,\rep2,1/6)$, U has the quantum numbers $(\rep3,\rep1,2/3)$, and D has the quantum numbers $(\rep3,\rep1,-1/3)$.}
\begin{center}
\begin{tabular}{|c|c|c|c|}  \hline
 $SU(2)_L$   & $U(1)_Y$  & $|Q|=|T_3+Y|$  &couplings to   \\ \hline
 $\rep1$         & $1/3$     & $1/3$         & QQ, UD         \\ \hline
 $\rep3$         & $1/3$     & $1/3,2/3,4/3$& QQ            \\ \hline
 $\rep1$         & $2/3$     & $2/3$         & DD            \\ \hline
 $\rep1$         & $4/3$     & $4/3$         & UU            \\ \hline
\end{tabular}
\end{center}
\label{qnum.TAB}
\end{table}

Any states with nontrivial $SU(3)_C$ color quantum numbers will interact with gluons and can thus be produced 
via gauge interactions. However, to understand the properties of exotic colored states, we must study their
model-dependent aspects. 
We consider a model parameterization in which, after electroweak symmetry breaking, a colored scalar 
diquark (denoted by $D$ henceforth) couples to quarks via the Lagrangian
\begin{equation}\label{eq:lagrangian}
\mathcal L=2\sqrt2\left[\bar K_i{}^{ab}D^i\bar q_a\big(\lambda_LP_L+\lambda_RP_R\big)q^C_b+\text{h.c.}\right],
\end{equation}
where $P_{L,R}\equiv(1\mp\gamma_5)/2$ are the left- and right-chirality projection operators, $q^C$ is the conjugate quark field, and the sum over quark flavors has been suppressed.
The scalar $D^i$ transforms according to either the sextet or antitriplet representation of $SU(3)_C$, and
the $\bar K_i{}^{ab}$ are Clebsch-Gordan coefficients coupling this representation to two triplets (see Appendix~\ref{app:color} for more details.)
Also, for different flavor combinations the couplings $\lambda_{L,R}$ may be completely independent, or they may be constrained by some underlying model.  Note that in the antitriplet case the couplings must be antisymmetric in flavor.

Existing data impose constraints on the couplings $\lambda$ in specific models.
For a review of constraints on $R$-parity-violating SUSY models see Ref.~\cite{Barbier:2004ez}.
In the case of a sextet diquark with a mass of a few hundred GeV to 1~TeV, coupling to right-handed up-type quarks, constraints arising from $D^0$-$\overline{D^0}$ mixing and the non-strange $D$ meson decay to $\pi\pi$
require \cite{Mohapatra:2007af}
\begin{equation}
\lambda_R^{uu},~\lambda_R^{uc}\lesssim0.1~{\rm and}~ \lambda_R^{cc}\approx0.
\end{equation}
Additionally, since the left-handed Cabibbo-Kobayashi-Maskawa (CKM) matrix is known, the left-handed couplings $\lambda_L$ are tightly constrained due to minimal flavor violation (MFV).
However, these constraints do not apply to the right-handed couplings $\lambda_R$, and our results are sensitive only to the combination $\lambda^2\equiv\lambda_L^2+\lambda_R^2$.
Since we do not assume any particular model, we do not concern ourselves with these constraints except to observe that Yukawa couplings of order 0.1 are allowed for the diquark masses of our interest.

\section{Leading-Order Scalar Production}
At leading order a scalar diquark is produced via the process
\begin{equation}
q(p_1)+q(p_2)\rightarrow D(l),
\end{equation}
where $p_1$, $p_2$, and $l$ are the momenta of the particles.
The invariant amplitude for this process is
\begin{equation}
\mathcal M^{(0)}=-2\sqrt2i\phi_iu^{\trans a}(p_1)K^i{}_{ab}(\lambda_LP_L+\lambda_RP_R)C^\dagger u^b(p_2),
\end{equation}
where $C$ is the charge conjugation matrix and $\phi_i$ is the color wavefunction of the diquark.
The Born-level partonic cross section is then
\begin{equation}\label{eq:Born}
\sigma_\text{Born}(\hat{s})=\frac{2\pi N_D{\lambda}^2}{N^2_C\hat s}\delta(1-\tau)\equiv\frac{\sigma_0}{\hat s}\delta(1-\tau),
\end{equation}
where $\hat{s}=(p_1+p_2)^2$, $\tau=m^2_D/\hat{s}$, $m_D$ is the mass of the diquark, $N_C=3$ is the number of colors in the fundamental representation, and $N_D$ is the dimension of the diquark's color representation.
That is, $N_D=6$ for a sextet diquark and $N_D=3$ for an antitriplet diquark.

The hadronic cross section is given by a convolution of the partonic cross section with parton distribution functions (pdfs):
\begin{equation}\label{eq:sigma_LO}
\sigma_\text{LO}=\frac{2\pi\lambda^2}S\frac{N_D}{N_C^2}(q\fold q)(\tau_0),
\end{equation}
where $S$ is the hadron collider's center-of-mass energy squared, $\tau_0\equiv m_D^2/S$, $q(x)$ is the quark distribution function (with a flavor sum again suppressed), and $\fold$ denotes the convolution, defined by
\begin{equation}
(f_1\fold f_2)(x)=\int_0^1dx_1\int_0^1dx_2\,\delta(x_1x_2-x)f_1(x_1)f_2(x_2),
\end{equation}
which is the $f_1 f_2$ parton luminosity. 

\section{Next-to-leading Order QCD Corrections}
Next-to-leading order QCD corrections to scalar diquark production arise through virtual gluon loops, the real-gluon emission process $qq\to gD$, and the gluon-initiated process $gq\to\bar qD$.
We give a general outline of the calculation in this section, and provide more details in Appendix~\ref{App:NLO}.

We do not consider the decay of the diquark.
Since the diquark carries color charge, a calculation taking its decay into account would include QCD corrections involving the exchange of color between the initial and final states.
Thus the NLO QCD corrections do not factorize perfectly into production and decay.
However, the effects of this imperfect factorization are suppressed by $\Gamma_D/m_D$, where $\Gamma_D$ is the decay width of the diquark.
In the present work we neglect such effects, and treat the diquark as a stable particle produced on shell.

\subsection{Virtual corrections}
\label{sec:nlov}
\begin{figure}[htb]
\begin{center}
\begin{picture}(60,100)(0,0)
\SetWidth{1.2}
\Vertex(35,50){1.3}
\ArrowLine(2,5)(35,50)
\ArrowLine(2,95)(35,50)
\DashArrowLine(35,50)(70,50)\dashlength
\Vertex(52,50){1.3}
\GlueArc(35,50)(25,0,124){3}{7}
\end{picture}~~~~
\begin{picture}(60,100)(0,0)
\SetWidth{1.2}
\Vertex(35,50){1.3}
\ArrowLine(2,5)(35,50)
\ArrowLine(2,95)(35,50)
\DashArrowLine(35,50)(70,50)\dashlength
\Vertex(52,50){1.3}
\GlueArc(35,50)(25,235,360){3}{7}
\end{picture}~~~~
\begin{picture}(60,100)(0,0)
\SetWidth{1.2}
\Vertex(35,50){1.3}
\ArrowLine(2,5)(35,50)
\ArrowLine(2,95)(35,50)
\DashArrowLine(35,50)(70,50)\dashlength
\Vertex(52,50){1.3}
\GlueArc(35,50)(25,124,236){3}{7}
\end{picture}~~~~
\begin{picture}(60,100)(0,0)
\SetWidth{1.2}
\Vertex(35,50){1.3}
\ArrowLine(2,5)(35,50)
\ArrowLine(2,95)(35,50)
\DashArrowLine(35,50)(70,50)\dashlength
\GlueArc(15,30)(12,36,248){2}{7}
\end{picture}~~~~
\begin{picture}(60,100)(0,0)
\SetWidth{1.2}
\Vertex(35,50){1.3}
\ArrowLine(2,5)(35,50)
\ArrowLine(2,95)(35,50)
\DashArrowLine(35,50)(70,50)\dashlength
\GlueArc(16,70)(12,108,326){2}{7}
\end{picture}~~~~
\begin{picture}(60,100)(0,0)
\SetWidth{1.2}
\Vertex(35,50){1.3}
\ArrowLine(2,5)(35,50)
\ArrowLine(2,95)(35,50)
\DashArrowLine(35,50)(70,50)\dashlength
\GlueArc(52,50)(12,0,180){2}{7}
\end{picture}~~~~
\caption{Feynman diagrams for virtual gluon corrections to $qq\to D$.}
 \label{fignlov}
\end{center}
\end{figure}
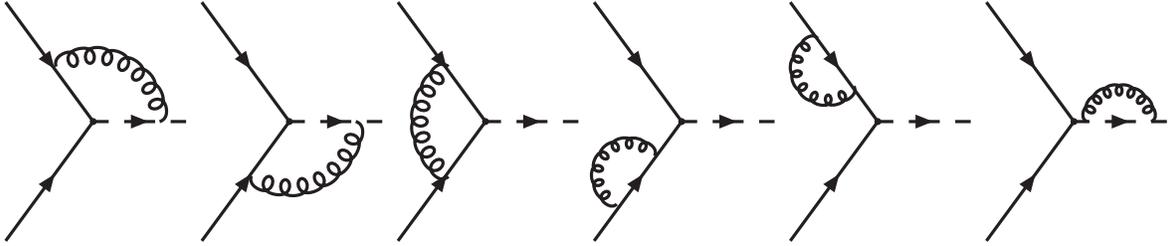

The one-loop diagrams contributing to the process $qq\to D$ are shown in Fig.~\ref{fignlov}.
These virtual-gluon loops give rise to ultraviolet (UV), soft, and collinear divergences.
All the divergences are regulated using dimensional regularization in $4-2\epsilon$ dimensions.

The UV divergences of the loop diagrams necessitate renormalization of the quark and diquark wavefunctions and of the $qqD$ vertex.
We perform this renormalization in the \MSbar\ scheme.
The wavefunction renormalization constants are found to be
\begin{equation}\label{eq:renorm-wf}
Z^q_2=1-\frac{\alpha_sC_F}{4\pi}\frac{(4\pi)^\epsilon}{\Gamma(1-\epsilon)}\frac{1}{\epsilon},\qquad Z^D_2=1+\frac{\alpha_s C_D}{2\pi}\frac{(4\pi)^{\epsilon}}{\Gamma(1-\epsilon)}\frac{1}{\epsilon},
\end{equation}
and the vertex renormalization constant is
\begin{equation}\label{eq:renorm-vtx}
Z_{\lambda}=1-\frac{\alpha_s}{4\pi}\frac{(4\pi)^\epsilon}{\Gamma(1-\epsilon)}\frac{1}{\epsilon}(4C_F-C_D).
\end{equation}
In these equations, $C_F$ and $C_D$ denote the eigenvalues of the quadratic Casimir operator of $SU(3)$ acting on the fundamental representation and on the diquark representation, respectively.
For a sextet diquark, $C_D=10/3$, and for an antitriplet diquark, $C_D=4/3=C_F$.  Since we are considering the single production of an on-shell diquark, the renormalization of the diquark mass is not needed for the calculation.

The effect of the self-energy diagrams and counterterms is to replace the coupling in the Born cross section by
\begin{equation}\label{eq:renorm-coup}
\lambda\to Z_\lambda(Z_2^q)^{-1}(Z_2^D)^{-1/2}\lambda,
\end{equation}
where $\lambda$ on the right-hand side denotes the renormalized coupling.
Once this replacement is performed, the remaining virtual corrections are given by the triangle diagrams in Fig.~\ref{fignlov}.

For the complete NLO QCD treatment, we also include the running coupling $\lambda(\mu_R^2)$, given by
\begin{equation}
\lambda(\mu_R^2)=\frac{\lambda(Q^2)}{1+\frac{3\alpha_s}{4\pi}\ln(\mu_R^2/Q^2)},
\label{eq:couprun}
\end{equation}
where $\mu_R$ is the renormalization scale.
\subsection{Real gluon emission}

\begin{figure}[htb]
\begin{center}
\begin{picture}(100,100)(0,0)
\SetWidth{1.2}
\Vertex(35,50){1.3}
\ArrowLine(2,20)(35,50)
\ArrowLine(2,80)(35,50)
\DashArrowLine(35,50)(65,50)\dashlength
\Gluon(20,60)(60,70){3}{5}
\end{picture}~~~~~
\begin{picture}(100,100)(0,0)
\SetWidth{1.2}
\Vertex(35,50){1.3}
\ArrowLine(2,20)(35,50)
\ArrowLine(2,80)(35,50)
\DashArrowLine(35,50)(65,50)\dashlength
\Gluon(20,35)(60,30){3}{5}
\end{picture}
\begin{picture}(100,100)(0,0)
\SetWidth{1.2}
\Vertex(35,50){1.3}
\ArrowLine(2,20)(35,50)
\ArrowLine(2,80)(35,50)
\DashArrowLine(35,50)(65,50)\dashlength
\Gluon(45,50)(65,70){3}{4}
\end{picture} \\
\caption{Feynman diagrams for $qq\to gD$}
\label{fignloqq}
\end{center}
\end{figure}
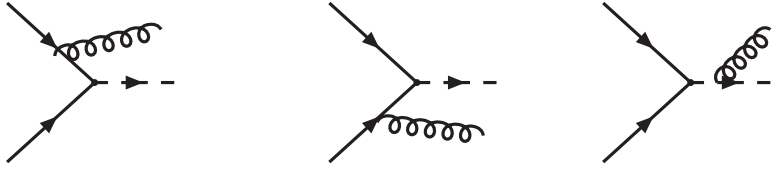
We must also consider the NLO corrections due to the radiation of an additional parton into the final state.
The process $qq\to gD$, shown in Fig.~\ref{fignloqq}, features both soft and collinear divergences.
The soft divergences cancel against those from the virtual correction, and the collinear divergences are absorbed into the pdfs using the \MSbar\ factorization scheme.

The total hadronic cross section corresponding to the partonic subprocess $qq\to D+X$ is then given at NLO by
\begin{equation}
\begin{split}\label{eq:qqtot}
\sigma_\text{NLO}^{qq}=&\frac{2\pi\lambda^2}S\frac{N_D}{N_C^2}\int_{\tau_0}^1\frac{d\tau}\tau\,(q\fold q)\left(\frac{\tau_0}\tau\right)\bigg[\delta(1-\tau)\\
&+\frac{\alpha_s}{2\pi}\bigg\{2P_{qq}(\tau)\ln\frac{m_D^2}{\mu_F^2\tau}\\
&+2C_F\left[2(1+\tau^2)\left(\frac{\ln(1-\tau)}{1-\tau}\right)_++\left(\frac{\pi^2}3-\frac32\ln\frac{m_D^2}{\mu_R^2}-1\right)\delta(1-\tau)+1-\tau\right]\\
&-C_D\left[\frac{1+\tau^2}{(1-\tau)_+}+\left(\frac{2\pi^2}3-1\right)\delta(1-\tau)\right]\bigg\}\bigg],
\end{split}
\end{equation}
where $\mu_F$ is the factorization scale, and $P_{qq}(\tau)=C_F\big[(1+\tau)^2/(1-\tau)\big]_+$ is the Dokshitzer-Gribov-Lipatov-Altarelli-Parisi (DGLAP) splitting function \cite{Altarelli:1977zs}.

\subsection{Gluon-initiated process}
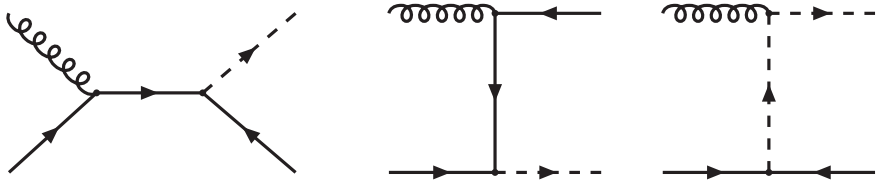
\begin{figure}[htb]
\begin{center}
\begin{picture}(100,100)(0,0)
\SetWidth{1.2}
\Vertex(35,50){1.3}
\Vertex(75,50){1.3}
\ArrowLine(2,20)(35,50)
\Gluon(35,50)(2,80){3}{5}
\ArrowLine(35,50)(75,50)
\ArrowLine(110,20)(75,50)
\DashArrowLine(75,50)(110,80)\dashlength
\end{picture}~~~~~~
\begin{picture}(100,100)(0,0)
\SetWidth{1.2}
\Vertex(60,80){1.3}
\Vertex(60,20){1.3}
\ArrowLine(20,20)(60,20)
\Gluon(20,80)(60,80){3}{5}
\ArrowLine(60,80)(60,20)
\ArrowLine(100,80)(60,80)
\DashArrowLine(60,20)(100,20)\dashlength
\end{picture}
\begin{picture}(100,100)(0,0)
\SetWidth{1.2}
\Vertex(60,80){1.3}
\Vertex(60,20){1.3}
\ArrowLine(20,20)(60,20)
\Gluon(20,80)(60,80){3}{5}
\DashArrowLine(60,20)(60,80)\dashlength
\DashArrowLine(60,80)(100,80)\dashlength
\ArrowLine(100,20)(60,20)
\end{picture} \\
\caption{Feynman diagrams for $gq\to\bar qD$.}
 \label{fignloqg}
\end{center}
\end{figure}
The $gq$ partonic channel contributes for the first time at NLO, via the process $gq\to\bar qD$, shown in Fig.~\ref{fignloqg}.
This process has only collinear divergences, which again are treated via the \MSbar\ factorization scheme.
The hadronic cross section corresponding to the $gq$ partonic channel is
\begin{multline}\label{eq:gqtot}
\sigma_\text{NLO}^{gq}=\frac{\lambda^2\alpha_s}S\frac{N_D}{N_C^2}\int_{\tau_0}^1\frac{d\tau}\tau\,(g\fold q+q\fold g)\left(\frac{\tau_0}\tau\right)\\
\times\left\{P_{qg}(\tau)\ln\frac{m_D^2(1-\tau)^2}{\mu_F^2\tau}-\frac14(1-\tau)(3-7\tau)+\frac{C_D}{C_F}\left[\tau\ln\tau+\frac12(1-\tau)(1+2\tau)\right]\right\},
\end{multline}
where $P_{qg}(\tau)=\frac12\big[\tau^2+(1-\tau)^2\big]$ is the DGLAP splitting function.

The total cross section for inclusive scalar diquark production is given by the sum of Eqs.~(\ref{eq:qqtot}) and (\ref{eq:gqtot}).

\begin{figure}[htb]
\centering
\subfigure[]{
	\includegraphics[width=0.35\textwidth,angle=-90]{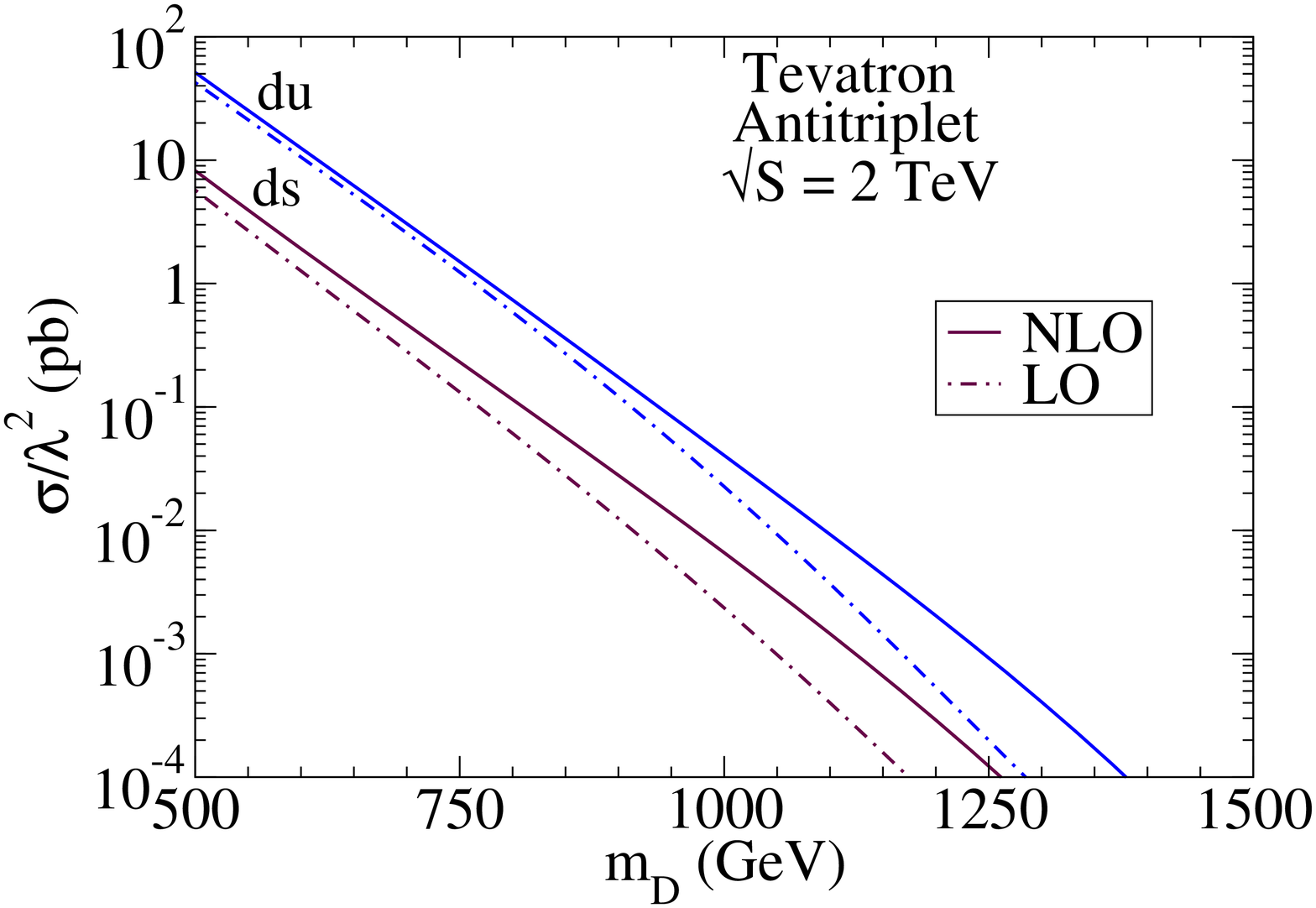}
	\label{fig:tevatrxsect}
	}
\subfigure[]{
	\label{fig:tevastxsect}
	\includegraphics[width=0.35\textwidth,angle=-90]{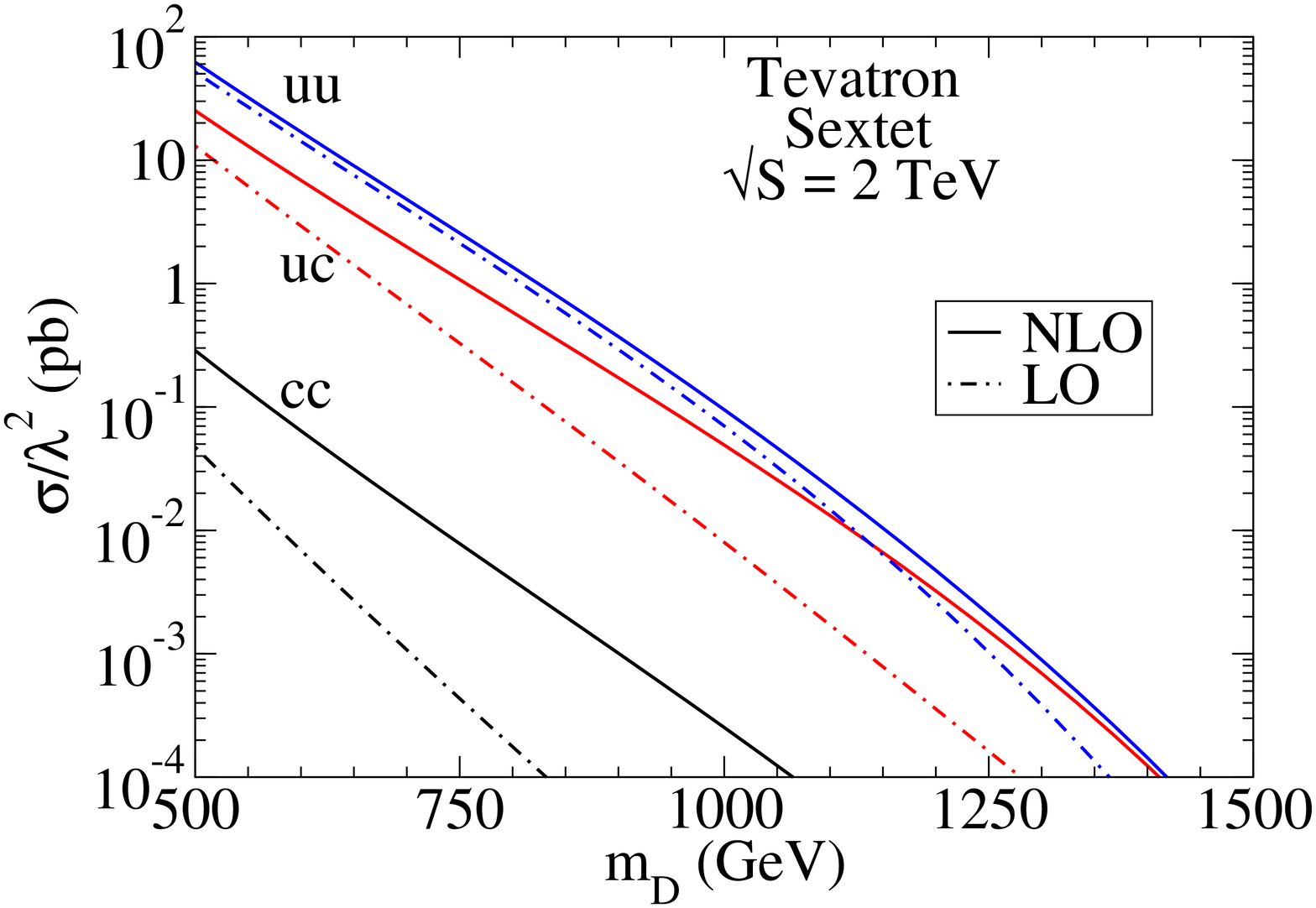}
	}
\subfigure[]{
        \label{fig:lhc10trxsect}
        \includegraphics[width=0.35\textwidth,angle=-90]{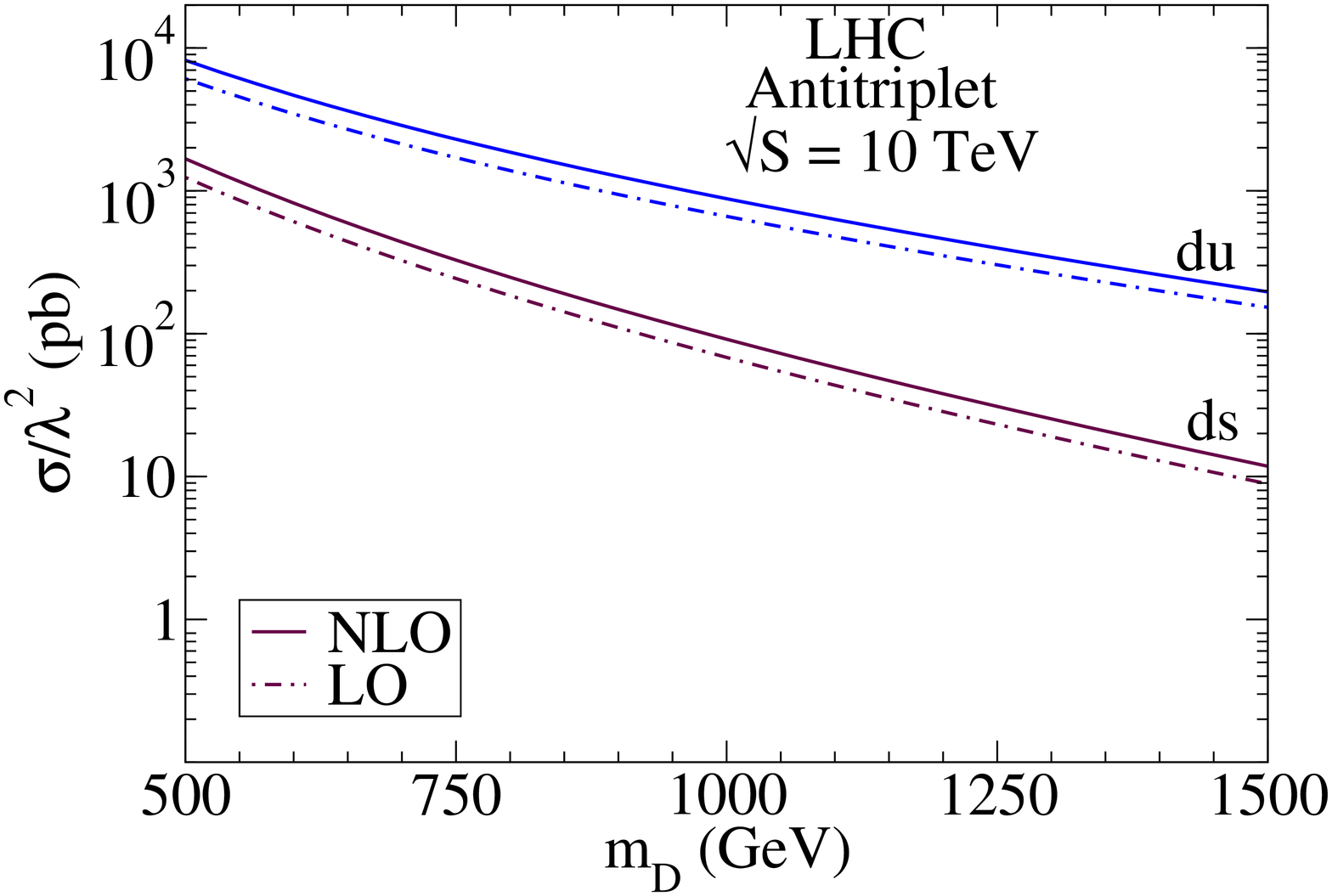}
        }
\subfigure[]{
        \label{fig:lhc10stxsect}
        \includegraphics[width=0.35\textwidth,angle=-90]{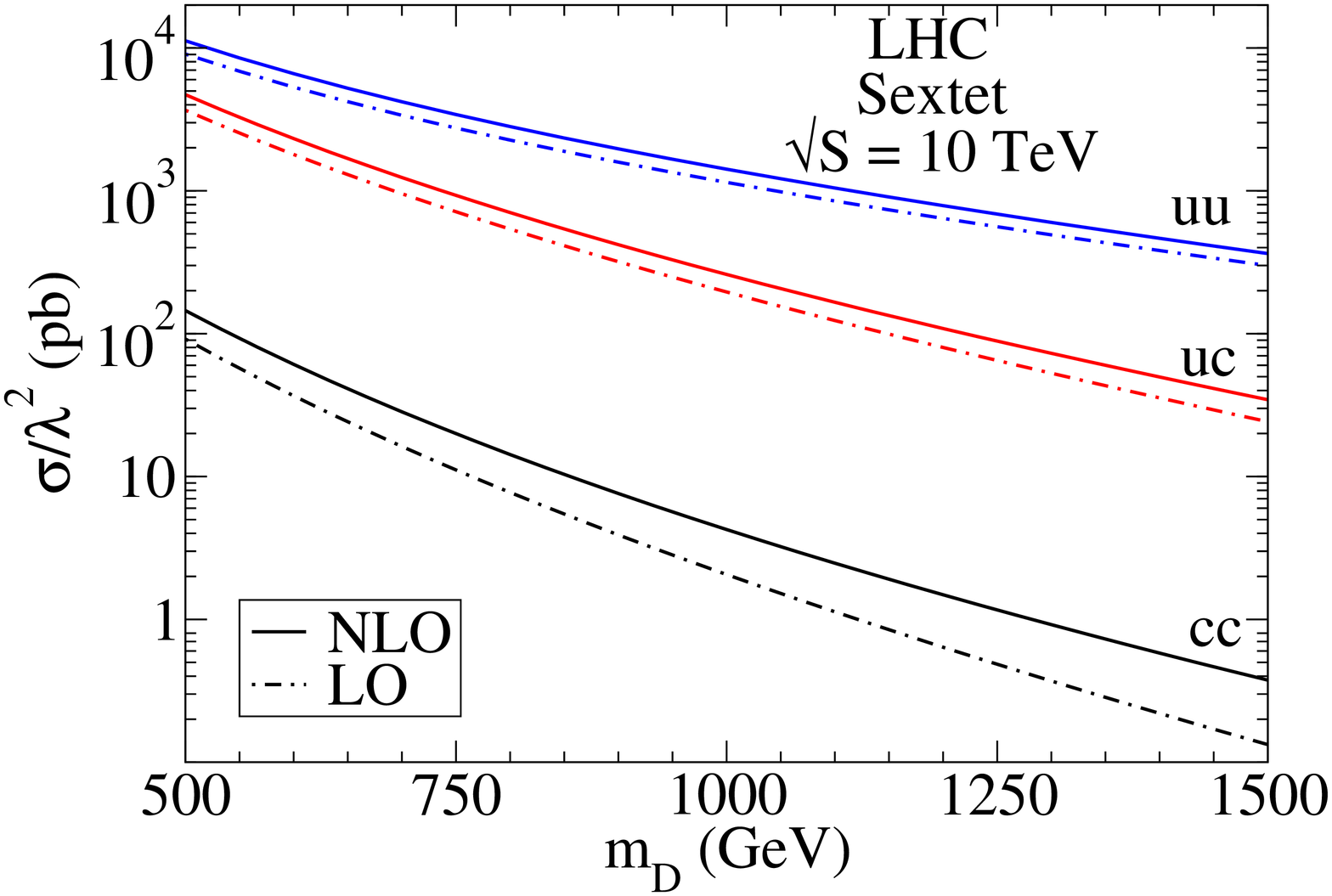}
        }
\caption{Results for $p\overline{p}$ collisions at a center of mass energy of $2$ TeV (a,b) and $pp$ collisions at 
10 TeV (c,d) with various initial states.  The total leading-order (dot-dash) and next-to-leading order (solid) cross sections are shown for both (a,c) the antitriplet and (b,d) the sextet diquarks.  For all the above plots the factorization scale, renormalization scale, and diquark mass are set equal.}
\par
\label{fig:xsect}
\end{figure}

\section{Numerical  Results}

In this section we present numerical results for the total cross section for production of a scalar diquark, for various diquark masses, center-of-mass energies, and initial states.
We use the CTEQ6L set of parton distribution functions \cite{Pumplin:2002vw} for leading-order results, and the CTEQ6.1M set \cite{Stump:2003yu} for NLO results.
Motivated by $R$-parity violating SUSY \cite{Barbier:2004ez}, for the antitriplet diquark we only consider initial states that can result in a final state with charge $-2/3$ or $+1/3$.  Also, following a partially unified Pati-Salam model~\cite{Mohapatra:2007af}, 
for the sextet case we only consider initial states that can result in a charge $+4/3$ final state.  Note that in the antitriplet case there are no identical initial partons since the couplings must be antisymmetric in flavor.  Unless otherwise noted, the factorization scale, renormalization scale, and diquark mass are set equal. 

Due to the much larger parton luminosity for the  valence quark initial states at the LHC, we only consider $qq$ contributions.
The inclusion of the $\bar{q}\bar{q}$ initial state would double the diquark production cross section for all initial states at the Tevatron (a $p\bar p$ collider) and for pure sea-quark initial states at the LHC.   
If a valence quark is in the initial state at the LHC, the $\bar{q}\bar{q}$ initial state contribution is much smaller.  For example, at the LHC at $14$~TeV with a diquark mass between $500$~GeV and $1.5$~TeV, the $uu$ initial state has a cross section 10--80 times larger than the $\bar{u}\bar{u}$ initial state.
\par
\begin{figure}[htb]
\centering
\subfigure[]{
	\includegraphics[width=0.35\textwidth,angle=-90]{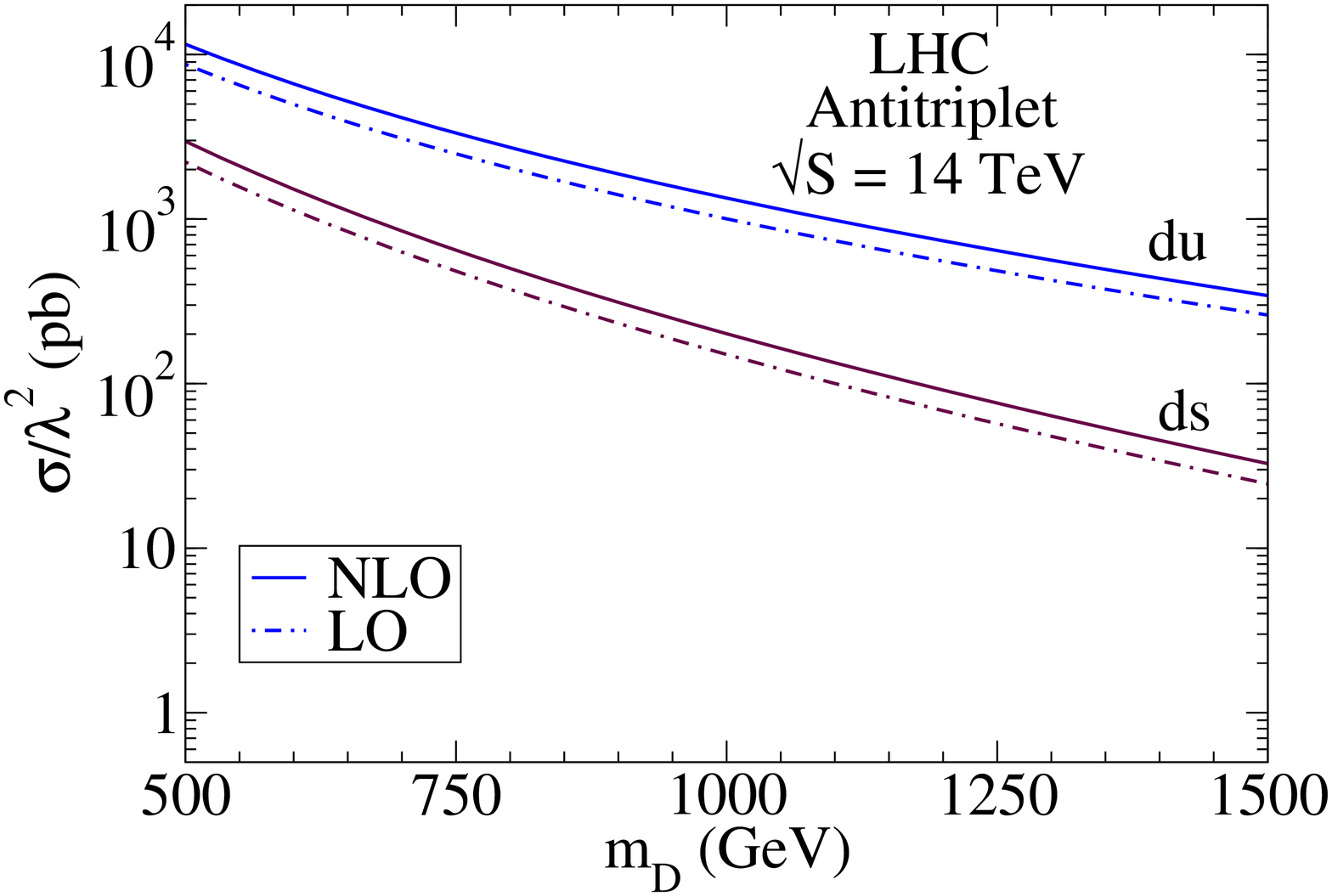}
	\label{fig:tripxsect}
	}
\subfigure[]{
	\label{fig:sextxsect}
	\includegraphics[width=0.35\textwidth,angle=-90]{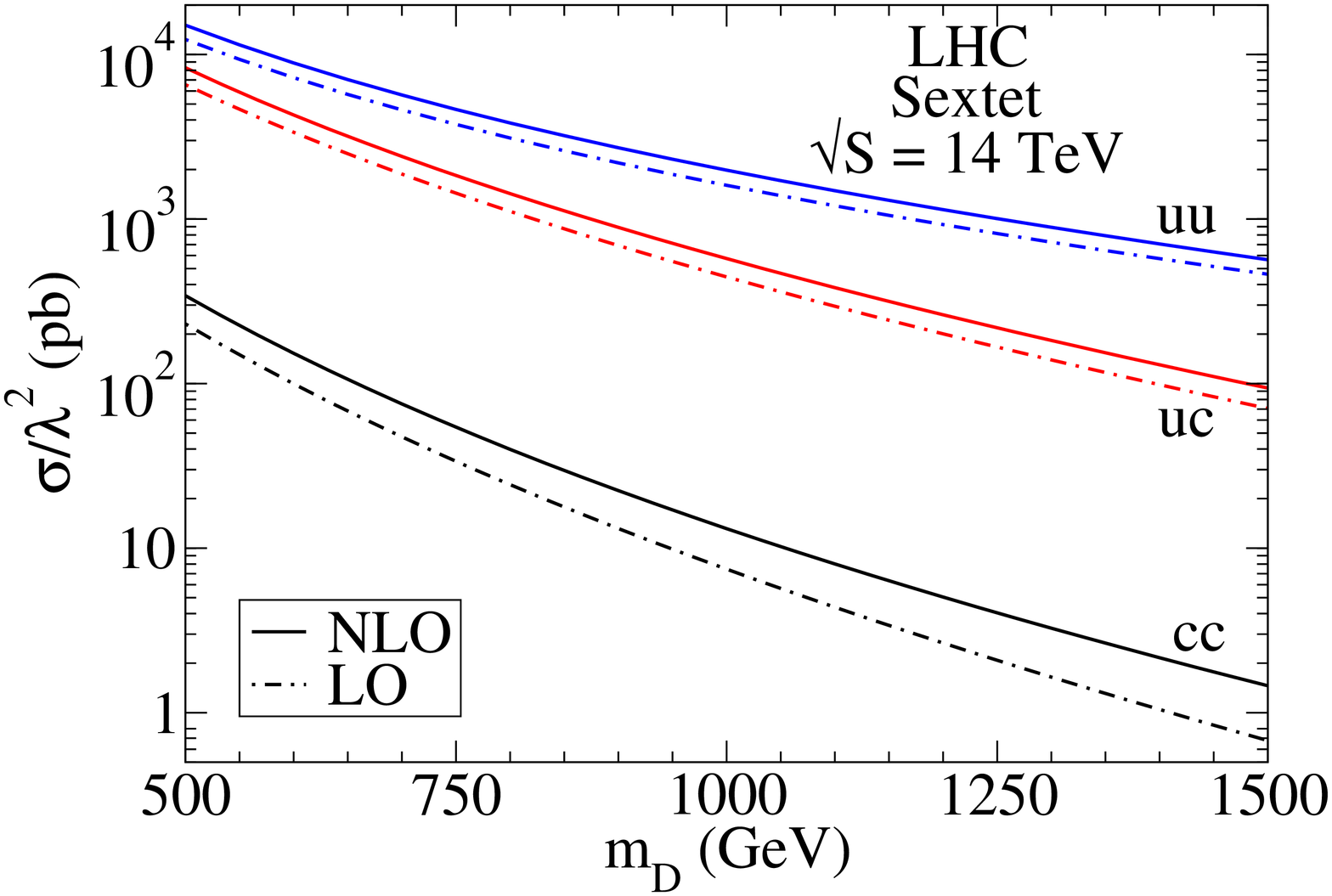}
	}
\subfigure[]{
        \label{fig:tripkfact}
        \includegraphics[width=0.35\textwidth,angle=-90]{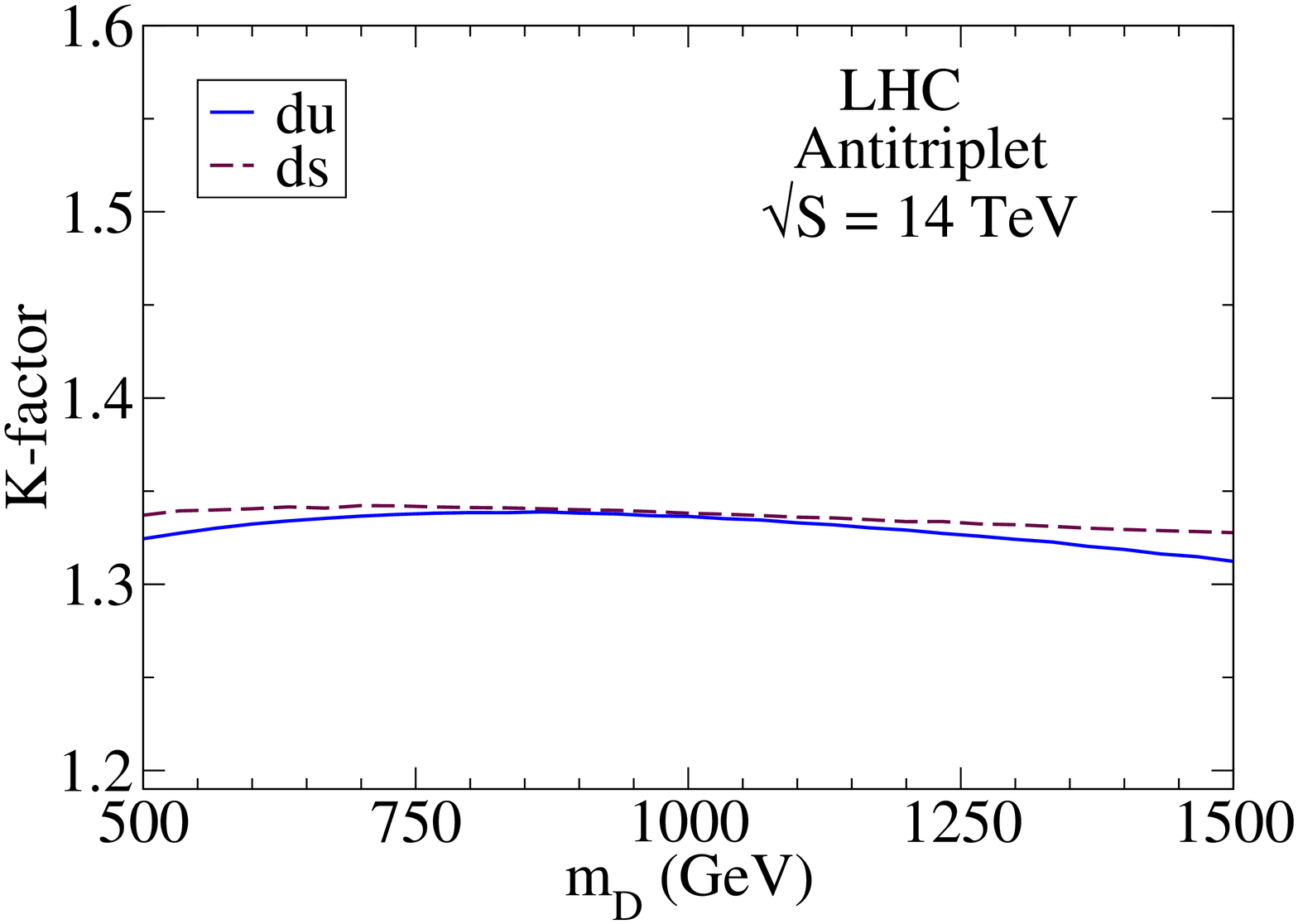}
        }
\subfigure[]{
        \label{fig:sextkfact}
        \includegraphics[width=0.35\textwidth,angle=-90]{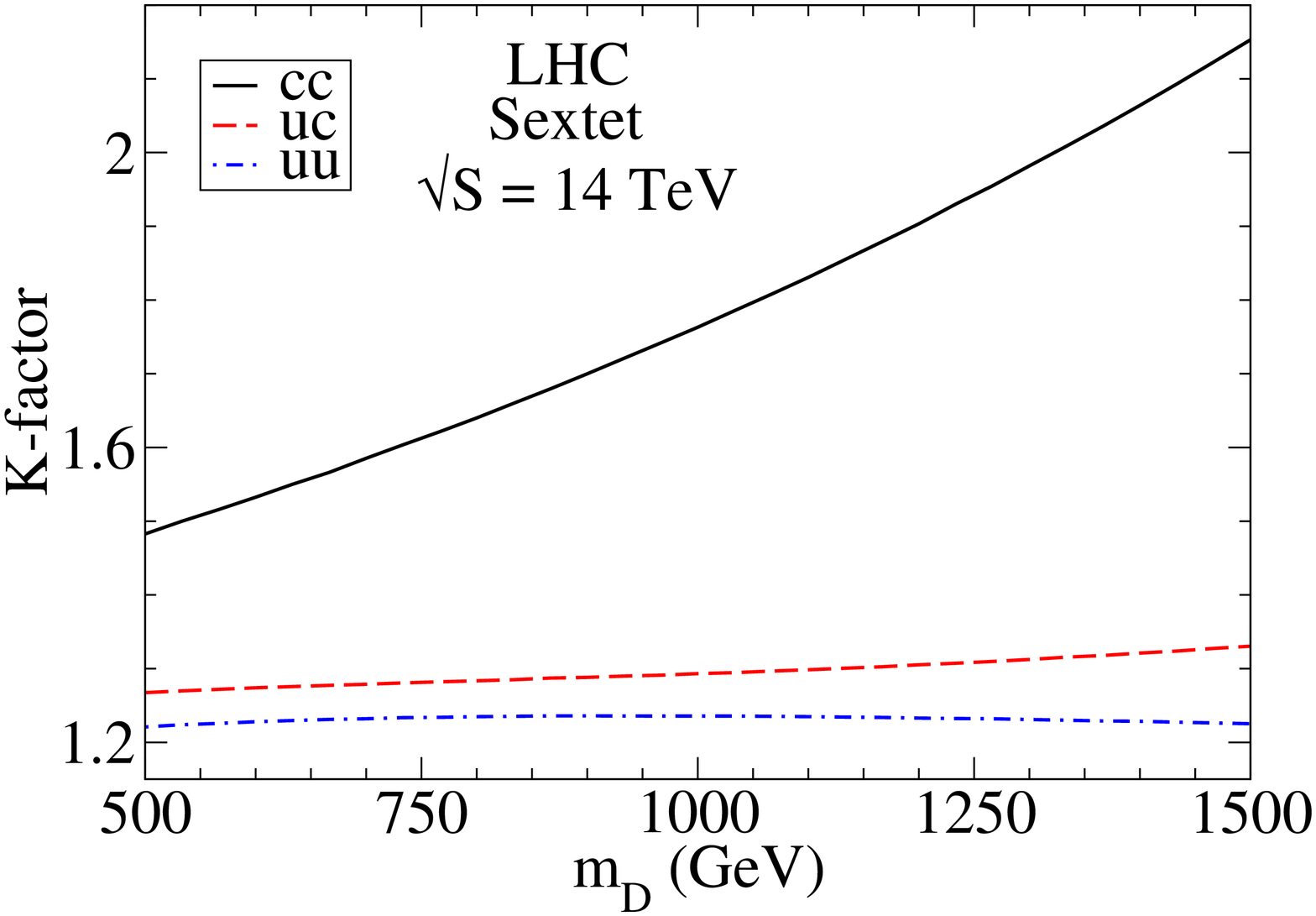}
        }
\caption{Results for $pp$ collisions at a center of mass energy of 14 TeV with various initial states.  The total leading-order (dot-dash) and next-to-leading order (solid) cross sections are shown for both (a) the antitriplet and (b) the sextet diquarks with various initial states.  Also shown are the $K$-factors for (c) the antitriplet and (d) the sextet diquarks with various initial states.  For all the above plots the factorization scale, renormalization scale, and diquark mass are set equal.}
\par

\label{fig:xsectkfact}
\end{figure}

The leading-order and NLO total cross section results for the antitriplet (sextet) diquark are shown in 
Fig.~\ref{fig:tevatrxsect} (Fig.~\ref{fig:tevastxsect}) for the Tevatron at $2$ TeV, Fig.~\ref{fig:lhc10trxsect} 
(Fig.~\ref{fig:lhc10stxsect}) for the LHC at $10$ TeV, and Fig.~\ref{fig:tripxsect} (Fig.~\ref{fig:sextxsect}) 
for the LHC at $14$ TeV.   We have factored out the coupling $\lambda^2$ for a model-independent
presentation. 
At lower diquark masses and not too small coupling, the Tevatron would produce scalar diquarks at reasonable rates,
reaching a cross section of the order of picobarns near $m_D\approx 700$ GeV.  
At the LHC at $10$ TeV the production cross section drastically increases over the Tevatron by a factor of
$10^2$--$10^4$ for $m_D=500\text{--}1000\,\text{GeV}$.
The production cross section at the LHC at $14$~TeV increases slightly over the $10$ TeV cross section.  
Also, the NLO cross section increases over the leading-order cross section in all cases.  If the couplings between the quarks and diquarks are not too small, the diquark will be produced at favorable rates. 

One generally expects the cross section for production of a sextet diquark to be larger than that of an antitriplet, due to a larger color factor in Eq.~(\ref{eq:sigma_LO}).
Thus, for example, the cross section of the sextet production from $uc$ is larger than that of the antitriplet production from $ds$.
The comparison of $uu$ sextet production to $du$ antitriplet production is an interesting case.
Apart from the larger color factor, $uu$ receives a relative enhancement due to the larger up-quark pdf.
However, $du$ also receives an enhancement, due to the combinatorics of the initial state: the $du$ luminosity is $d\fold u +u\fold d$, while the $uu$ luminosity is simply $u\fold u$.
These luminosity enhancements approximately offset one another, so that the $uu$ cross section is approximately twice that of $du$.

\begin{figure}[htb]
\centering
\subfigure[]{
	\includegraphics[width=0.35\textwidth,angle=-90]{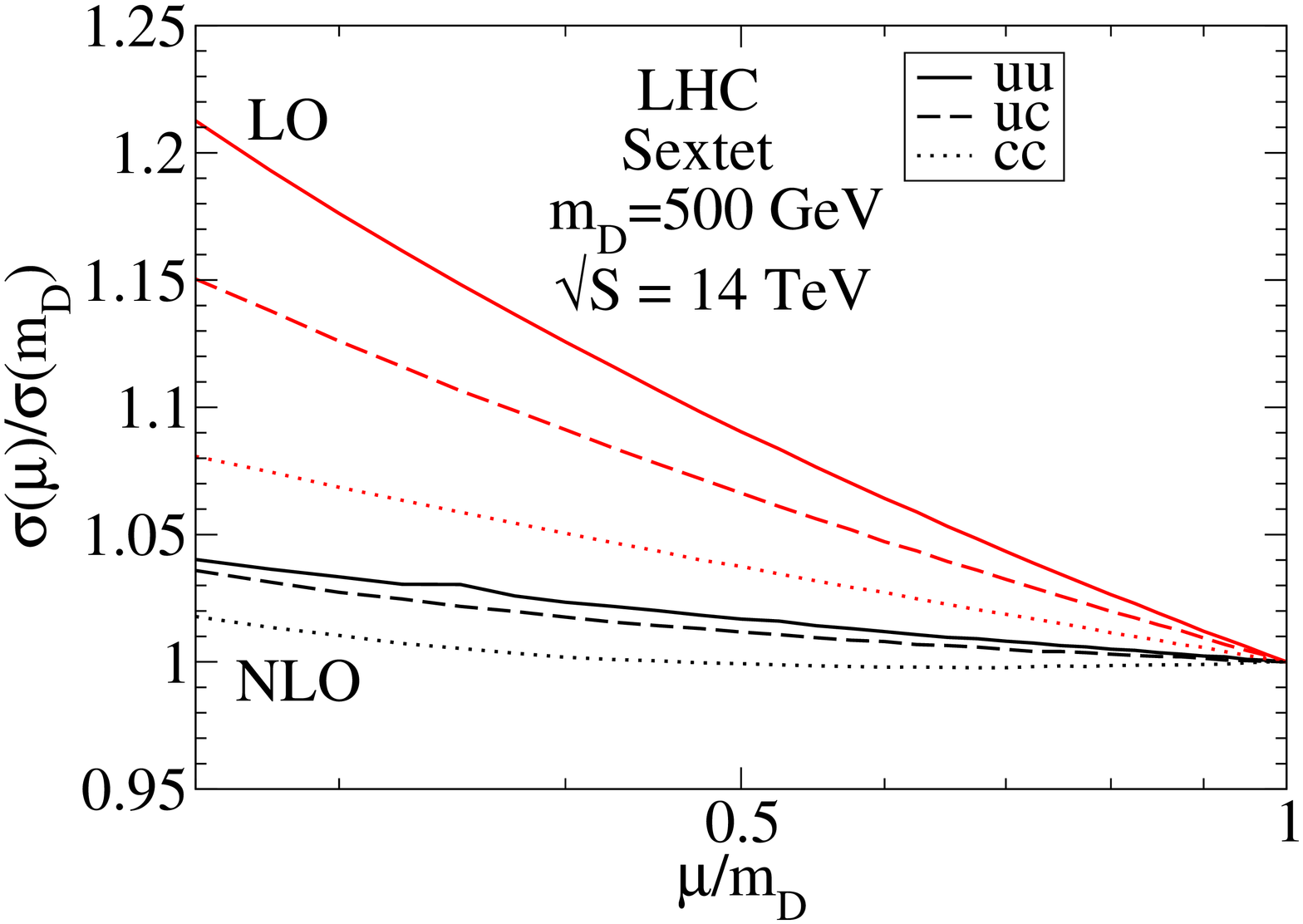}
	\label{fig:sextmu500}
	}
\subfigure[]{
	\label{fig:sextmu1000}
	\includegraphics[width=0.35\textwidth,angle=-90]{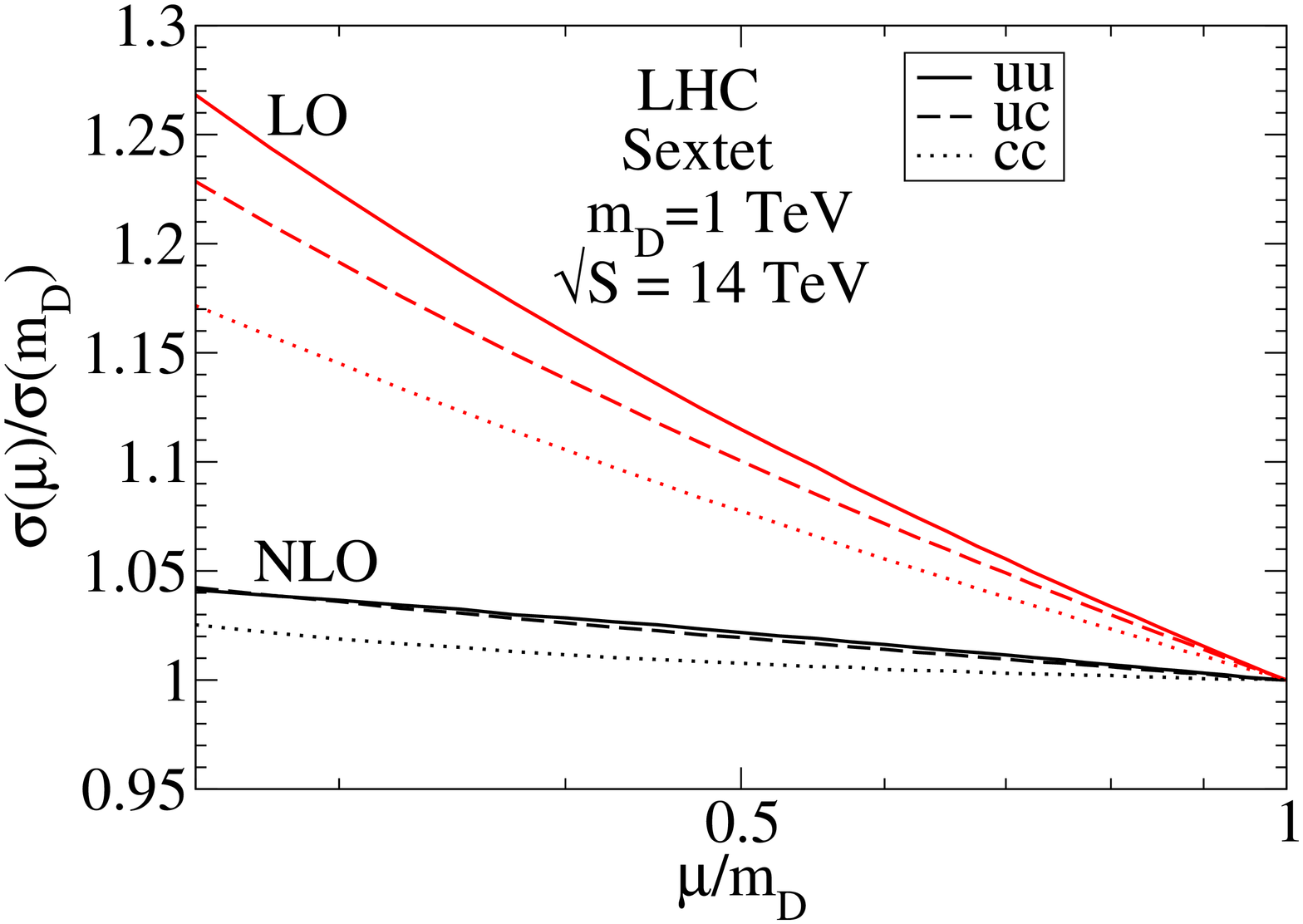}
	}
\subfigure[]{
        \label{fig:tripmu500}
        \includegraphics[width=0.35\textwidth,angle=-90]{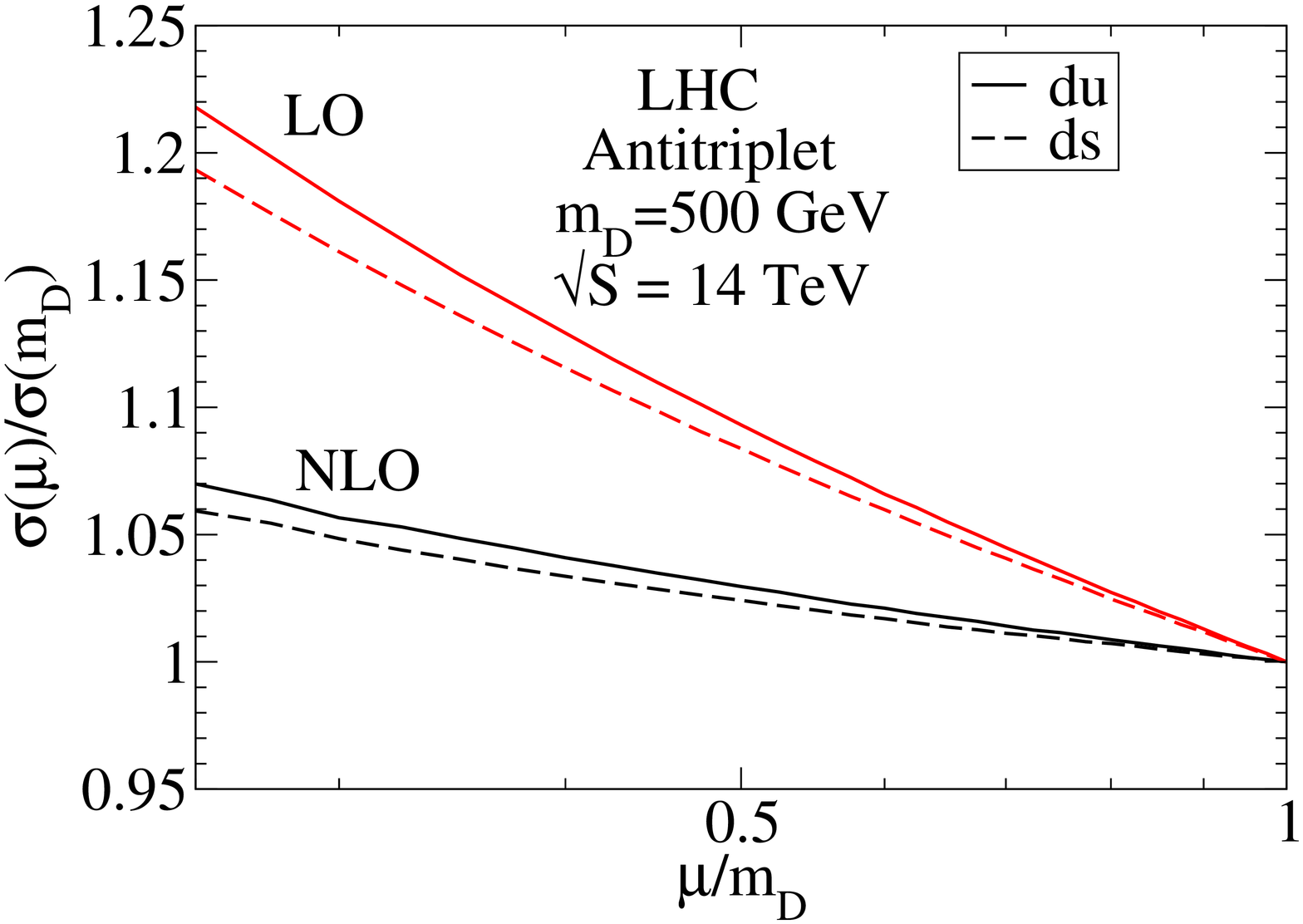}
        }
\subfigure[]{
        \label{fig:tripmu1000}
        \includegraphics[width=0.35\textwidth,angle=-90]{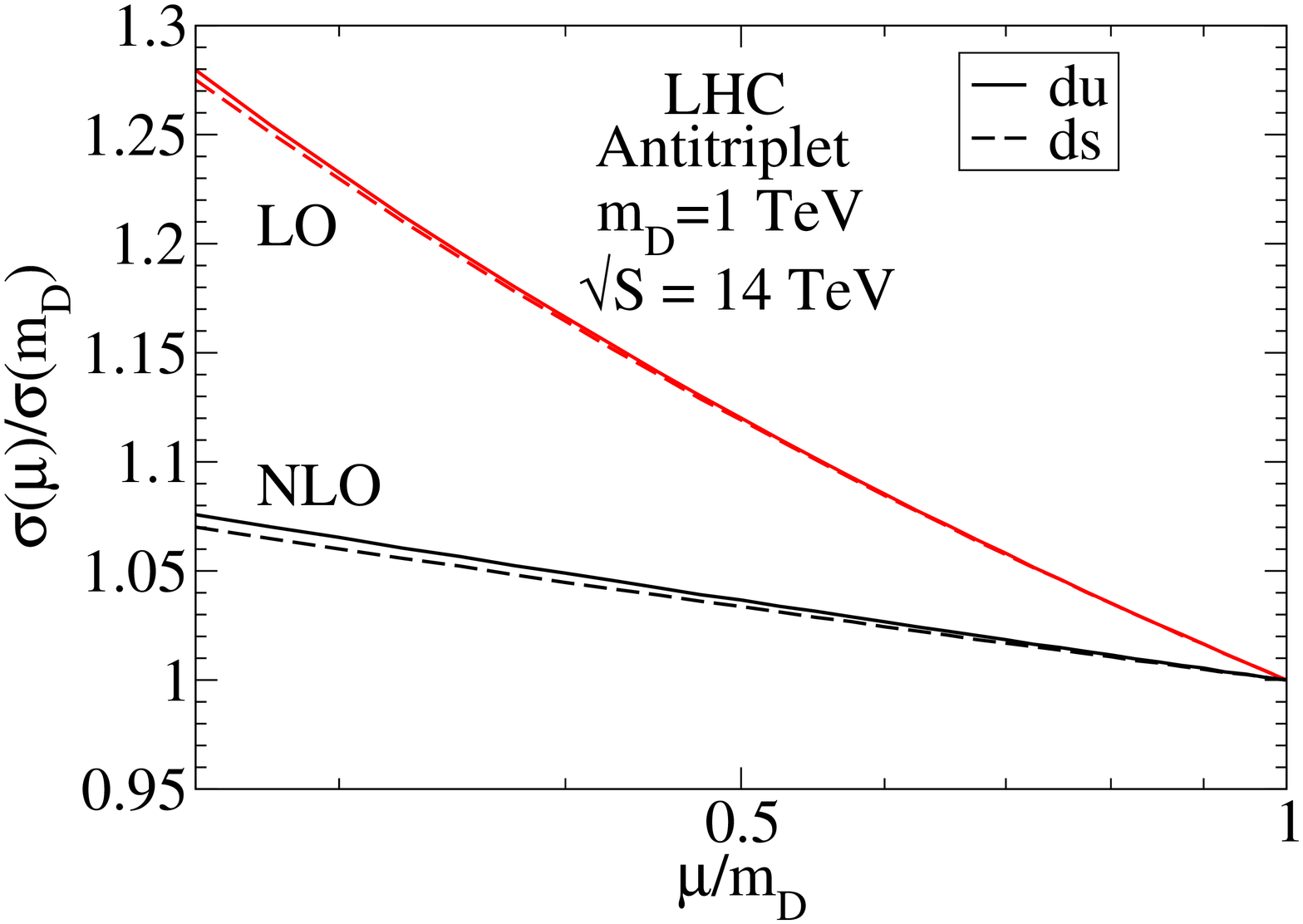}
        }
\caption{Leading-order (red) and next-to-leading order (black) factorization and renormalization scale dependence for $pp$ collisions at a center of mass energy of 14 TeV and various initial states.  Figures (a) and (b) show the factorization and renormalization scale dependence for a $500$~GeV and $1$~TeV sextet diquark, respectively, and figures (c) and (d) show the scale depence of a $500$~GeV and $1$~TeV antitriplet diquark, respectively. The factorization scale and renormalization scales are set equal in all plots.}
\par
\label{fig:scale}
\end{figure}

\par
An important quantity for NLO calculations is the $K$-factor, defined as the ratio of the NLO cross section to the leading-order cross section.
Figures \ref{fig:tripkfact}, \ref{fig:sextkfact} show the $K$-factor as a function of the diquark mass for the antitriplet and sextet cases, respectively. 
 In the antitriplet case the $K$-factor is between $1.31$ and $1.35$.
   For the sextet case, the $K$-factor runs from around $\sim1.27$ to $\sim1.32$ for the up-charm initial state, is around $1.22$ for the up-up initial state, and is between $\sim1.49$ and  $\sim2.15$ for the charm-charm initial state.  The purely sea-quark initial states ($cc$) show more dependence of the $K$-factor on the diquark mass because of the differences in the factorization-scale dependence between the leading-order and NLO pdfs.  For initial states involving up and down quarks, the $K$-factor for the sextet case is generally smaller than that of the antitriplet case.
This surprising result is due to a partial cancellation between the two color structures [the $C_F$ and $C_D$ terms in Eq.~(\ref{eq:qqtot})].

The factorization and renormalization scale dependence is shown in Figs.~\ref{fig:sextmu500}, \ref{fig:sextmu1000} for a $500$ GeV and a $1$ TeV sextet diquark, respectively, and in Figs.~\ref{fig:tripmu500}, \ref{fig:tripmu1000} for a $500$ GeV and a $1$ TeV antitriplet diquark, respectively.  For all the figures, the factorization and renormalization scales were set equal and varied from $m_D/4$ to $m_D$.
As expected, the scale dependence of the NLO cross section is less than that of the leading-order cross section for all cases.

We have also computed the rapidity distribution of a $1\,\text{TeV}$ sextet diquark at the LHC.
The shape of the NLO distribution is virtually identical to that of the leading-order distribution; the only significant difference is in the overall normalization.
Thus, for purposes of the rapidity distribution, the effect of the NLO corrections is well described by a simple $K$-factor.

In  Table \ref{xsect.TAB} we tabulate the cross sections at leading order, leading order with NLO pdfs, and full NLO, as well as the exclusive cross sections for production of a diquark with an extra parton (gluon or antiquark) in the final state.
This final-state parton is required to have a transverse momentum of at least $20\,\text{GeV}$.
When a parton appears in the final state, it is more often a gluon than an antiquark, by a ratio of about $3:1$ if $m_D=500\,\text{GeV}$, or about $8:1$ if $m_D=1\,\text{TeV}$.
This is due to the increasing dominance of the valence-quark parton luminosity over the gluon parton luminosity as the diquark mass increases.

\begin{table}[htb]
\caption{$14$ TeV LHC total cross section (in nanobarns) at leading order, leading order with NLO pdfs, full NLO, diquark plus a gluon, and diquark plus an antiquark.
A minimum transverse momentum of $20\,\text{GeV}$ is required for final-state partons.
For the sextet (antitriplet) diquark case the initial state is $uu$ ($du$).
In this table the coupling $\lambda$ is taken to be 1; the cross sections scale as $\lambda^2$.}
\renewcommand{\arraystretch}{1.2}
\begin{center}
\begin{tabular}{|c|c|r|r|r|r|r|}  \hline
Color & $m_D$~[TeV] & $\sigma^\text{CTEQ6L}_\text{LO}$ [nb] & $\sigma^\text{CTEQ6.1M}_\text{LO}$ & $\sigma^\text{CTEQ6.1M}_\text{NLO}$ & $\sigma^\text{CTEQ6.1M}_{D+g}$ & $\sigma^\text{CTEQ6.1M}_{D+\bar q}$ \\ \hline
$\rep6$ & 0.5         & $12.3$   & $13.3$    & $15.1$    & $4.07$           & $1.58$\\ \hline
$\rep6$ & 1           & $1.60$           & $1.72$            & $1.98$       & $0.785$        & $0.101$\\ \hline
$\rep{\bar3}$ & 0.5         & $8.71$ & $9.47$   & $11.5$     & $3.48$       & $1.05$\\ \hline
$\rep{\bar3}$ & 1           & $1.00$ & $1.08$   & $1.34$     & $0.573$        & $0.0673$\\ \hline
\end{tabular}
\end{center}
\label{xsect.TAB}
\end{table}

As for the possible observation of the diquark signal, 
the decay of the diquark to two light-quark jets may be masked by SM QCD dijet events.  
The decay to two tops can be more distinctive, in particular for the like-sign charge combinations of electrons
and muons from the $tt$ decay, although 
the branching fractions of the diquark to two tops may not be too large and will be model-dependent \cite{Mohapatra:2007af}.

\section{Soft Gluon Resummation and the Transverse Momentum Distribution}

In Table \ref{xsect.TAB}, we listed some cross section values for a diquark plus a final-state parton. These rates are very sensitive 
to the cutoff on the parton transverse momentum $p_T$. 
For transverse momenta above other scales in the process, say the diquark mass, fixed order perturbation theory in $\alpha_s$ gives reliable results.  If the transverse momentum is much smaller than the diquark mass, the series in $\alpha_s$ is replaced by a series in $\alpha_s\ln^2(m^2_D/p^2_T)$~\cite{Parisi:1979se,Collins:1981uk,Collins:1981va}. As $p_T\rightarrow 0$ the expansion becomes non-perturbative, but it is possible to resum the terms that are at least as divergent as $p^{-1}_T$.
Using the results in the previous section, we can obtain the full $p_T$ distribution of the diquark at the leading log to all
orders in $\alpha_s$. 

For the production of strongly interacting particles, the formalism for the resummation of small transverse momentum logarithms has not been fully developed~\cite{Bozzi:2005wk}.  Although it has not been proven, we assume that the final-state and initial-state radiation can be resummed in the same way.  This assumption has been made previously for $t\bar t$ production at the Tevatron \cite{Mrenna:1996cz} and scalar top quark production at hadron colliders \cite{Plehn:2000be}.

Following the usual procedure \cite{Arnold:1990yk}, the leading-order asymptotic cross section in the low $p_T$ limit is
\begin{eqnarray}
\frac{d^2\sigma^\text{asym}}{dp_T\,dy}&&=\sigma_0\frac{\alpha_s}{\pi}\frac{1}{Sp_T}\bigg{[}\bigg{(}A\ln\frac{m^2_D}{p^2_T}+B\bigg{)}f_q(x^0_1)f_{q'}(x^0_2)+(P_{qq}\otimes f_q)(x^0_1)f_{q'}(x^0_2)\\
&&+f_q(x^0_1)(P_{qq}\otimes f_{q'})(x^0_2)+(P_{qg}\otimes f_g)(x^0_1)f_{q'}(x^0_2)+f_q(x^0_1)(P_{qg}\otimes f_{q'})(x^0_2)\bigg{]}+(x^0_1\leftrightarrow x^0_2),\nonumber
\end{eqnarray}
where $y$ is the diquark rapidity, $x^0_{1,2}=(m_D/\sqrt{S})e^{\pm y}$ are the dominant terms at low $p_T$, and the coefficients $A,B$ are found to be $A=2C_F$ and $B=-(3C_F+C_D)$.  Our results for $A$ and $B$ agree with the previous result for single top squark production~\cite{Plehn:2000be}.  The resummed formula for diquark production is

\begin{eqnarray}
\frac{d^2\sigma^\text{resum}}{dp_T\,dy}&=&\sigma_0\frac{2p_T}{S}\int^\infty_0db\,\frac{b}{2}J_0(bp_T)W(b), \label{eq:Wb}\\
W(b)&=&\exp\bigg{[}-\int^{m^2_D}_{b^2_0/b^2}\frac{dq^2}{q^2}\frac{\alpha_s(q^2)}{2\pi}\bigg{(}A\ln\frac{m^2_D}{q^2}+B\bigg{)}\bigg{]}f_q(x_1^0)f_{q'}(x_2^0)+\bigg{(}x^0_1\leftrightarrow x^0_2\bigg{)},\nonumber
\end{eqnarray}
where $J_0(x)$ is a zeroth-order Bessel function, $W(b)$ is a Sudakov form factor from the sum of large logs in impact parameter space, the canonical value of $b_0$ is $2e^{-\gamma_E}$, and the pdfs are evaluated at a factorization scale of $\mu_F=b_0/b$.  If the running of $\alpha_s$ is kept to leading order, the integral in Eq.~(\ref{eq:Wb}) can be evaluated.  
Also, $\Lambda_{\rm QCD}$ was set so that the leading-order $\alpha_s(M^2_Z)$ attained the correct value.

\par
 For $b\ge1/\Lambda_{QCD}$ confinement sets in and $\alpha_s$ diverges.  To cut off the divergence and parameterize the non-perturbative effects, we make the replacement \cite{Collins:1984kg,Davies:1984hs,Davies:1984sp}
\begin{eqnarray}
W(b)\rightarrow W(b_*)\exp\bigg{[}-b^2g_1-b^2g_2\ln\frac{b_\text{max} m_D}{2}\bigg{]},\qquad b_*=\frac{b}{\sqrt{1+b^2/b^2_\text{max}}}, 
\end{eqnarray}
where $g_1=0.14\,\text{GeV}^2$, $g_2=0.54\,\text{GeV}^2$, and $b_\text{max}=(2\,\text{GeV})^{-1}$ \cite{Landry:2002ix}.  There are other methods of parameterizing the non-perturbative effects \cite{Landry:2002ix}, but the results did not significantly differ from this parameterization.

The coefficient $A=2C_F$ is of the usual form found from resummation of the Drell-Yan production~\cite{Davies:1984hs,Davies:1984sp}, while the other coefficient $B=-(3C_F+C_D)$ contains the usual term $-3C_F$ found in the resummation of Drell-Yan production~\cite{Davies:1984hs,Davies:1984sp} and a new term proportional to $C_D$.
In order to investigate the sensitivity of the results to the assumption that initial- and final-state radiation resum in the same way, we have also computed the resummed distribution without final-state radiation, \textit{i.e.}~with $B=-3C_F$.
We find that this causes the peak in the resummed distribution for the antitriplet $du$ (sextet $uu$) initial state to decrease by about $20\%$ ($45\%$).
Also, the peak in the resummed distrbution for the $500$~GeV ($1$~TeV) diquark shifts to the right by about $2$ GeV ($1$~GeV).
Since $C_D$ is larger in the sextet case than in the antitriplet case, setting $B=-3C_F$ has a larger effect on the resummed distribution in the sextet case.

The resummed distribution has only accounted for the terms in the perturbative expansion that are at least as divergent as $p^{-1}_T$.  In the high-$p_T$ region other terms become important and the resummed distribution is inaccurate.  To match between the perturbative and non-perturbative regions the total $p_T$ distribution can be defined as 
\begin{eqnarray}
\frac{d^2\sigma^\text{total}}{dp_T\,dy}=\frac{d^2\sigma^\text{pert}}{dp_T\,dy}+f(p_T)\bigg{(}\frac{d^2\sigma^\text{resum}}{dp_T\,dy}-\frac{d^2\sigma^\text{asym}}{dp_T\,dy}\bigg{)},\quad f(p_T)=\frac{1}{1+(p_T/p^{\text{match}}_T)^4},
\end{eqnarray}
where $p^{\text{match}}_T$ is a scale above which the the perturbative distribution is accurate, and $f(p_T)$ is a matching function.  The function $f$ is needed, since at large $p_T$ the resummed and asymptotic distributions are invalid and their difference can be larger than the perturbative distribution's value.  The large power of $p_T$ in the denominator of $f(p_T)$ is chosen so that the matching function quickly goes to zero as $p_T$ increases above $p^{\rm match}_T$, and quickly goes to one as $p_T$ goes to zero.

\begin{figure}[htb]
\centering
\subfigure[]{
	\includegraphics[width=0.36\textwidth,angle=-90]{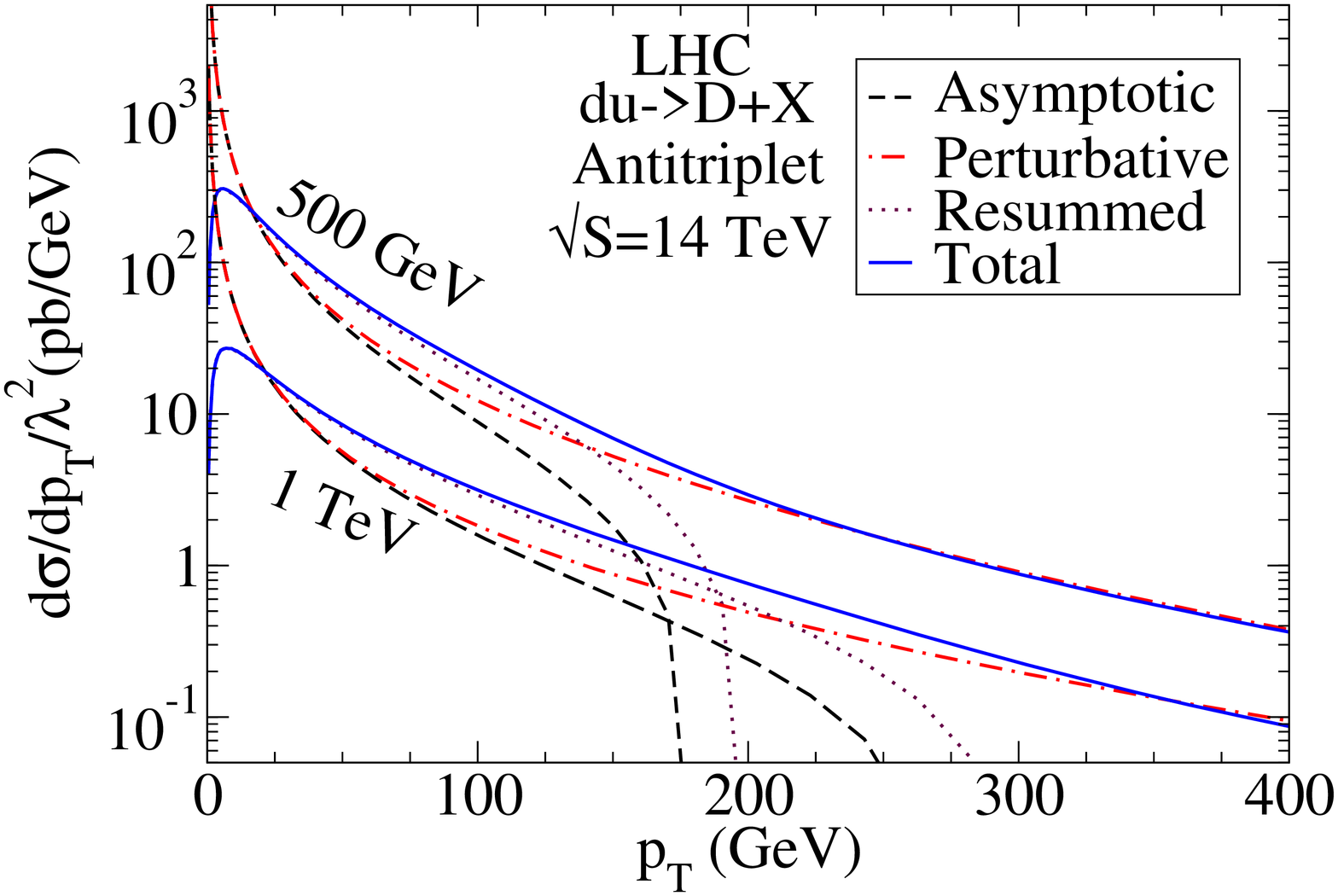}
	\label{fig:ptdutrip}
	}
\subfigure[]{
	\includegraphics[width=0.36\textwidth,angle=-90]{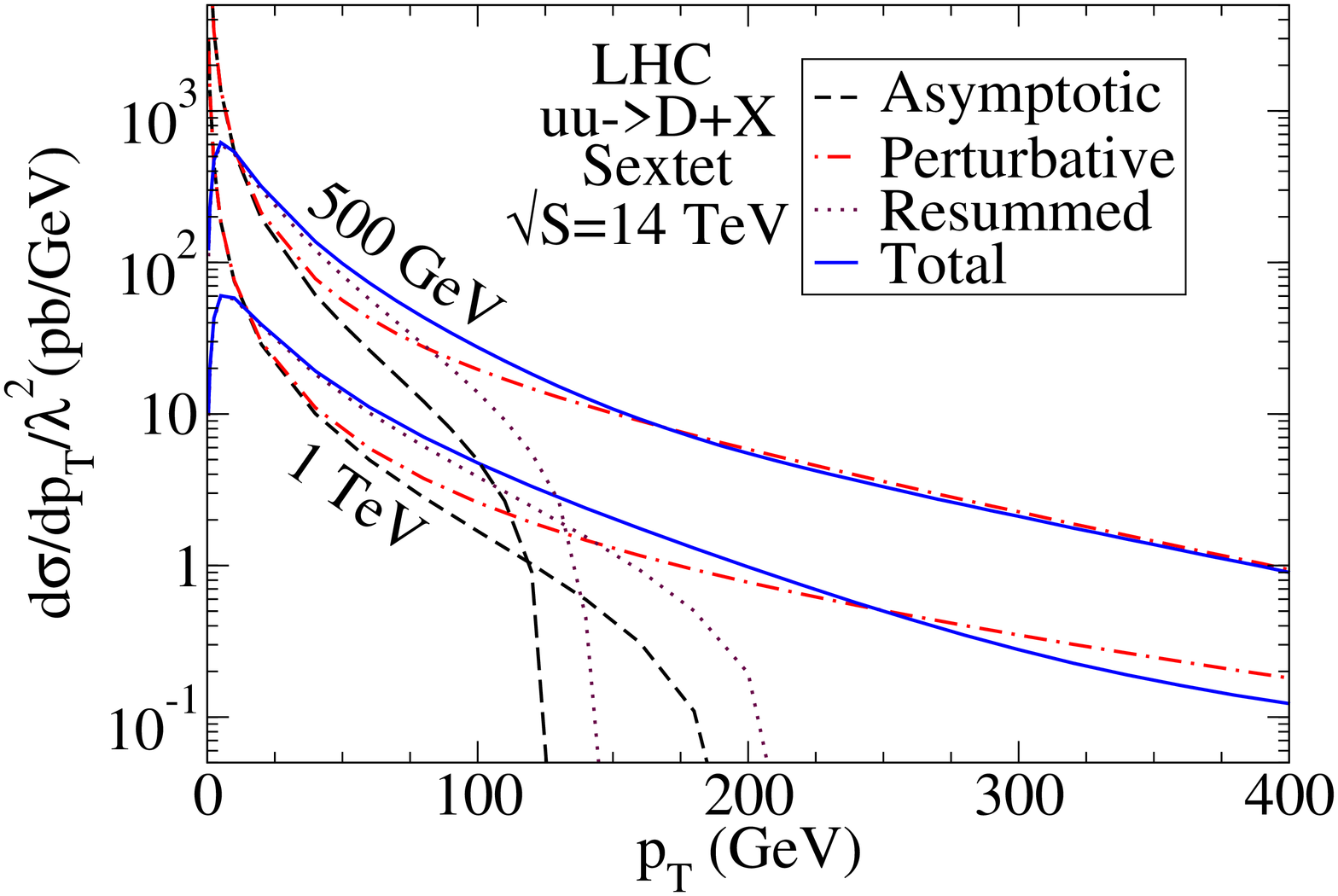}
	\label{fig:ptuusext}
	}
\caption{$p_T$ distributions or diquarks with masses of $500$ GeV and $1$ TeV.  (a) shows the $p_T$ distribution of an antitriplet diquark with $du$ initial state, and (b) shows the $p_T$ distribution of a sextet diquark with $uu$ initial state.  Both (a) and (b) show the aymptotic (dashed), perturbative (dot-dash), resummed (dotted), and total (solid) distributions. }
\label{fig:ptdist}
\end{figure}

Figure \ref{fig:ptdist} shows the $p_T$ distributions for both a $500$ GeV and $1$ TeV diquark.  The antitriplet $p_T$ distribution with down-up initial state is shown in Fig.~\ref{fig:ptdutrip}, and the sextet $p_T$ distribution with up-up initial state is shown 
in Fig.~\ref{fig:ptuusext}.  For all the distributions the typical value of $p^{\text{match}}_T=m_D/3$ was used.  At low $p_T$ the perturbative and asymptotic distributions cancel and the total distribution matches the resummed distribution, while at high $p_T$ the matching function goes to zero and, beginning around $p_T=m_D/3$, the total distribution converges to the perturbative distribution.   As is typical, the low $p_T$ distributions peak around $5$--$8$ GeV.  Hence, the $p_T$ distribution no longer diverges as $p_T\rightarrow 0$ and the perturbative distribution becomes reliable around $m_D/3$.



\section{Summary}
The LHC is a hadron machine, hence any new particle participating in QCD interactions can be produced with favorable rates.  Due to the gluon's high parton luminosity, the LHC is typically refered to as a ``gluon factory.''  For rather heavy final states the gluon parton luminosity decreases relative to the valence-quark parton luminosity, hence valence-quark scattering is still important.  Quark-quark annihilation can result in colored sextet and antitriplet scalars, so-called diquarks.  

We calculated the NLO corrections to the single production of these exotic colored states.  The production cross section at the LHC was favorable, provided that the coupling between the diquark and SM quarks is not too suppressed.  For most initial states and a diquark mass between $500$~GeV and $1.5$ TeV, the $K$-factor for the antitriplet diquark is between $\sim1.31$ and $\sim 1.35$, and for the sextet diquark the $K$-factor is around $1.22$ for the up-up initial state and runs from $\sim1.27$ to $\sim1.32$ for the up-charm initial state.   If the initial state is composed exclusively of sea quarks, the $K$-factors are much larger due to the differences between the leading-order and NLO pdfs.  Also, the NLO calculation was found to decrease the renormalization and factorization scale dependence for all initial states and both the antitriplet and sextet cases.

The soft gluon resummation for low transverse momentum was also calculated for the antitriplet diquark from the down-up initial state and the sextet diquark from the up-up initial state with diquark masses of $500$~GeV and $1$~TeV.  At low $p_T$ the total $p_T$ distribution matches the resummed distribution, peaking around $5$--$8$~GeV.  For $p_T\gtrsim m_D/3$ the total distribtution converges to the perturbative distribution, indicating that for $p_T$ above $m_D/3$ the perturbative distribution is reliable.
  

\section{Acknowledgement}
We would like to thank Z.G.~Si for discussions. 
This work was supported in part by the US DOE under contract No.~DE-FG02-95ER40896.

\appendix
\section{Color}
\label{app:color}
The product of two fundamental representations of $SU(3)$ is $\rep3\otimes\rep3=\rep6\oplus\rep{\bar3}$, so a boson produced in $qq$ fusion can be either a color sextet or an antitriplet.
Since the sextet especially is rather unfamiliar, we include in this Appendix some technology for working out the color factors involved in our calculation.
While the gauge group $SU(3)$ is of primary interest, we present results for arbitrary $SU(N_C)$.

\subsection{The fundamental representation}
We denote a vector transforming under the fundamental representation of $SU(N_C)$ by a raised Latin index: $u^a$.
An object with a lowered index, $\bar u_a$, belongs to the antifundamental.
The generators of $SU(N_C)$ in the fundamental representation are denoted by $t^{Aa}{}_b$, or just $t^A$ in matrix notation.
Uppercase Latin indices belong to the adjoint representation; they can be freely raised and lowered using the Kronecker symbols $\delta^{AB}$ and $\delta_{AB}$.
The generators are given by $t^A=\frac12\lambda^A$, where the $\lambda^A$'s are the familiar Gell-Mann matrices, and they satisfy the commutation relation
\begin{equation}
[t^A,t^B]=if^{ABC}t_C,
\end{equation}
where $f^{ABC}$ are the structure constants of $SU(N_C)$.
The quadratic Casimir operator is given by
\begin{equation}
t^At_A=C_F=\frac{N_C^2-1}{2N_C},
\end{equation}
and the generators satisfy the orthogonality relation
\begin{equation}
\Tr t^At^B=T_F\delta^{AB}=\frac12\delta^{AB},
\end{equation}
and the completeness relation
\begin{equation}\label{eq:t-complete}
t^{Aa}{}_bt_A{}^c{}_d=\frac12\left(\delta^a_d\delta^c_b-\frac1{N_C}\delta^a_b\delta^c_d\right).
\end{equation}

\subsection{Clebsch-Gordan coefficents}
The generators $\mathbf T^A$ in the product of two fundamental representations can be written as
\begin{equation}\label{eq:prod-rep}
\mathbf T^{Aab}{}_{cd}=t^{Aa}{}_c\delta^b_d+\delta^a_ct^{Ab}{}_d.
\end{equation}
By a unitary change of basis we can bring $\mathbf T^A$ into block-diagonal form and read off the irreducible representations.
There are always two irreducible representations; the symmetric (antisymmetric) combination of two fundamentals has dimension $N_D=N_C(N_C\pm1)/2$.%
\footnote{Whenever the symbols $\pm$ and $\mp$ are used in this Appendix, the upper (lower) sign refers to the symmetric (antisymmetric) combination of two fundamentals.}
Suppose we are interested in only one of these representations.
We can assign it raised mid-alphabet latin indices, and from the change-of-basis matrix we can read off the Clebsch-Gordan coefficients $K^i{}_{ab}$.
We can choose explicit values for the sextet of $SU(3)$:
\begin{equation}
\begin{split}
&K^1=\begin{pmatrix}1&0&0\\0&0&0\\0&0&0\end{pmatrix},\qquad
K^2=\frac1{\sqrt2}\begin{pmatrix}0&1&0\\1&0&0\\0&0&0\end{pmatrix},\qquad
K^3=\begin{pmatrix}0&0&0\\0&1&0\\0&0&0\end{pmatrix},\\
&K^4=\frac1{\sqrt2}\begin{pmatrix}0&0&0\\0&0&1\\0&1&0\end{pmatrix},\qquad
K^5=\begin{pmatrix}0&0&0\\0&0&0\\0&0&1\end{pmatrix},\qquad
K^6=\frac1{\sqrt2}\begin{pmatrix}0&0&1\\0&0&0\\1&0&0\end{pmatrix},
\end{split}
\end{equation}
and for the antitriplet:
\begin{equation}
K^1=\frac1{\sqrt2}\begin{pmatrix}0&0&0\\0&0&1\\0&-1&0\end{pmatrix},\qquad
K^2=\frac1{\sqrt2}\begin{pmatrix}0&0&-1\\0&0&0\\1&0&0\end{pmatrix},\qquad
K^3=\frac1{\sqrt2}\begin{pmatrix}0&1&0\\-1&0&0\\0&0&0\end{pmatrix}.
\end{equation}
Note that, in the special case of $SU(3)$, the antisymmetric combination of two fundamentals is an antifundamental; thus we can replace the raised index $^i$ with a lowered index $_a$ and write $K_{abc}=\epsilon_{abc}/\sqrt2$.
It will also be useful to define Clebsch-Gordan coefficients for the conjugate representation: $\bar K_i{}^{ab}\equiv(K^i{}_{ba})^*=K^i{}_{ba}$ since we can always choose real coefficients.
The coefficients $K^i{}_{ab}$ are normalized such that
\begin{equation}
K^i\bar K_i=N_D/N_C=(N_C\pm1)/2,
\end{equation}
and they satisfy the orthogonality relation
\begin{equation}
\Tr K^i\bar K_j=\delta^i_j,
\end{equation}
and the completeness relation
\begin{equation}\label{eq:K-complete}
K^i{}_{ab}\bar K_i{}^{cd}=\frac12(\delta_a^d\delta_b^c\pm\delta_a^c\delta_b^d).
\end{equation}

\subsection{Generators in the diquark representation}
We can form the generators $T^A$ in the diquark representation simply by extracting the appropriate block from the product representation:
\begin{equation}
T^{Ai}{}_j=K^i{}_{ab}\bar K_j{}^{dc}\mathbf T^{Aab}{}_{cd}.
\end{equation}
Applying Eq.~(\ref{eq:prod-rep}) and simplifying, we find
\begin{equation}\label{eq:T-def}
T^{Ai}{}_j=2\Tr K^it^A\bar K_j.
\end{equation}
Using this definition, it is straightforward to verify that the $T^A$ satisfy the appropriate commutation relation:
\begin{equation}
[T^A,T^B]=if^{ABC}T_C.
\end{equation}

The quadratic Casimir operator in the diquark representation is given by
\begin{equation}
T^AT_A=C_D=(N_C\mp1)(N_C\pm2)/N_C.
\end{equation}
The generators $T^A$ also satisfy the orthogonality relation
\begin{equation}
\Tr T^AT^B=T_D\delta^{AB}=\frac12(N_C\pm2)\delta^{AB},
\end{equation}
and together with their products $T^AT^B$ they satisfy the completeness relation
\begin{equation}\label{eq:T-complete}
(T^AT^B)^i{}_j(T_AT_B)^k{}_l\mp\frac{N_C\mp4}{N_C}T^{Ai}{}_jT_A{}^k{}_l-\frac{N_C^2\pm2N_C-4}{N_C^2}\delta^i_j\delta^k_l=\delta^i_l\delta_j^k.
\end{equation}
Due to its greater complexity, Eq.~(\ref{eq:T-complete}) is not as useful as its counterparts Eqs.~(\ref{eq:t-complete}) and (\ref{eq:K-complete}).
Nevertheless, we include it here for the sake of completeness.

\subsection{Color factors}
Provided we sum or average over the color states of external particles,  
color matrices in the calculation of a cross section always appear with all indices contracted.
Therefore it is straightforward to evaluate any color factor using Eqs.~(\ref{eq:t-complete}), (\ref{eq:K-complete}), and (\ref{eq:T-def}).
The color factors appearing in our calculation are:
\begin{equation}
\Tr K^i\bar K_i=N_D=\frac{N_C(N_C\pm1)}2,
\end{equation}
\begin{equation}
\Tr K^it^At_A\bar K_i=C_FN_D=\frac{(N_C^2-1)(N_C\pm1)}4,
\end{equation}
\begin{equation}
T^{Ai}{}_jT_A{}^j{}_k\Tr K^k\bar K_i=C_DN_D=\frac{(N_C^2-1)(N_C\pm2)}2,
\end{equation}
\begin{equation}
\Tr K^it^A\bar K_i(t_A)^\trans=\pm\frac12C_FN_C=\pm\frac{N_C^2-1}4,
\end{equation}
\begin{equation}
T^{Ai}{}_j\Tr K^jt_A\bar K_i=\frac12C_DN_D=\frac{(N_C^2-1)(N_C\pm2)}4,
\end{equation}

\section{Details of the NLO Calculation}
\label{App:NLO}
Here we give the details of the calculation of the NLO corrections to single diquark production.  All divergences are regulated using dimensional regularization in $4-2\epsilon$ dimensions and the \MSbar\ scheme is used to cancel the ultraviolet and collinear divergences.
\subsection{Virtual corrections}
The Feynman diagrams contributing to the virtual NLO corrections are shown in Fig.~\ref{fignlov}.  The quark self-energy diagrams do not contribute since the quarks have no mass.  Also, in the Feynman gauge, the contribution from the diquark self-energy diagram is found to be zero.
Hence, as mentioned in Section~\ref{sec:nlov}, the self-energy diagrams and counterterms modify the Born cross section via the replacement \cite{Peskin:1995ev}
\begin{equation}
\lambda\rightarrow Z_{\lambda}(Z^q_2)^{-1}(Z^D_2)^{-1/2}\lambda,
\end{equation}
where $Z^q_2$ and $Z^D_2$ are the wavefunction renormalization constants of the quark and diquark, respectively, $Z_{\lambda}$ is the vertex renormalization constant, and the coupling constant on the right-hand side is renormalized.  The expressions for the renormalization constants are given in Eqs.~(\ref{eq:renorm-wf}) and (\ref{eq:renorm-vtx}).  All the following coupling constants are understood to be renormalized.

The remaining virtual correction comes from the triangle diagrams.
Performing the loop integrals, the invariant amplitude for the vertex corrections is
\begin{eqnarray}
\mathcal{M}_\text{vertex}&=&
-2\sqrt{2}i\phi_{i} u^\trans(p_1)C^\dagger(\lambda_R P_R+\lambda_L P_L)\frac{\alpha_s}{4\pi}C_\epsilon\bigg{[}t_A^\trans K^{i}{t^A}(\frac{2}{\epsilon^2}+2-\pi^2)\nonumber\\
&&+(t_A^\trans K^{j}+K^{j}{t_A}){T^{Ai}}_j\left(-\frac{1}{\epsilon^2}-\frac{1}{\epsilon}-2-\frac{\pi^2}{6}\right)\bigg{]}u(p_2),\\
C_{\epsilon}&=&\frac{1}{\Gamma(1-\epsilon)}\bigg{(}\frac{4\pi\mu^2_R}{\hat{s}}\bigg{)}^\epsilon,
\end{eqnarray}
where the color indices of the fundamental representation are suppressed and $\mu_R$ is the renormalization scale.
Using the identities from Appendix \ref{app:color}, the contribution to the cross section from vertex corrections is found to be
\begin{eqnarray}
\hat\sigma_\text{vertex}(\hat{s})&=&\frac{\alpha_s}{2\pi}C_{\epsilon}\bigg{[}2C_F\bigg{(}-\frac{1}{\epsilon^2}-1+\frac{\pi^2}{2}\bigg{)}-C_D\bigg{(}1+\frac{2\pi^2}{3}+\frac{1}{\epsilon}\bigg{)}\bigg{]}\sigma_\text{Born}(\hat{s})\nonumber\\
&\equiv& K^{'}\sigma_\text{Born}(\hat{s}).
\end{eqnarray}

The sum of the Born and virtual diagrams is then
\begin{eqnarray}
\sigma_\text{Born}(\hat s)+\hat\sigma_\text{vertex}(\hat s)&=&\frac{\sigma_0}{\hat s}\delta(1-\tau)\bigg{\{}\bigg{[}Z_{\lambda}(Z^q_2)^{-1}(Z^D_2)^{-1/2}\bigg{]}^2+K^{'}\bigg{\}}\nonumber\\
&=&\frac{\sigma_0}{\hat s}\delta(1-\tau)\bigg{\{}1-\frac{\alpha_s}{2\pi}C_{\epsilon}\bigg{[}C_D\bigg{(}\frac{1}{\epsilon}+1+\frac{2}{3}\pi^2\bigg{)}
\label{eq:vet}\\
&&+2C_F\bigg{(}\frac{1}{\epsilon^2}+\frac{3}{2\epsilon}+\frac{3}{2}\ln\frac{\hat s}{\mu^2_R}+1-\frac{\pi^2}{2}\bigg{)}\bigg{]}\bigg{\}},\nonumber
\end{eqnarray}
where $\sigma_0=2\pi N_D\lambda^2/N^2_C$.  The renormalization constants have canceled the UV divergences, hence the remaining $\epsilon$ poles are from collinear and soft divergences.

\subsection{Real gluon emission}
The process
\begin{eqnarray}
q(p_1)+q(p_2)\rightarrow g(k)+D(l),
\end{eqnarray}
as shown in Fig. \ref{fignloqq}, also contributes to NLO corrections.
The invariant amplitude for this process is
\begin{equation}
\mathcal{M}_{qq}=\begin{aligned}[t]
&-2\sqrt{2}i\mu^\epsilon_R g_s\epsilon^A_\mu(k)\phi_{i}{u}^\trans(p_1)C^\dagger (\lambda_R P_R+\lambda_L P_L)\\
&\times\bigg{[}K^{j}T_A{}^i_j\frac{(2l+k)^\mu}{\hat{s}-m^2_D}-t_{A}^\trans K^{i}\gamma^\mu\frac{\slashed{p}_1-\slashed{k}}{2p_1\cdot k}-K^{i}t_{A}\frac{\slashed{p}_2-\slashed{k}}{2p_2\cdot k}\gamma^\mu\bigg{]}u(p_2),
\end{aligned}
\end{equation}
where the quark color indices have been suppressed and the diquark color indices have been left explicit.  Squaring the amplitude and summing over color results in
\begin{eqnarray}
|\mathcal{M}_{qq}|^2=16{\lambda}^2g^2_sN_D\mu^{2\epsilon}_R\bigg{(}\frac{2\tau}{(1-\tau)^2}+1-\epsilon\bigg{)}\bigg{(}C_F\frac{4}{\sin^2\theta}-C_D\bigg{)},
\end{eqnarray}
where  $\theta$ is the angle between the gluon and one of the initial-state quarks.
\par
The collinear and soft divergences of real emission are regulated by using the $4-2\epsilon$ dimensional Lorentz-invariant phase space:
\begin{eqnarray}
\int dPS^\epsilon_2&=&\int \frac{d^{3-2\epsilon}k}{(2\pi)^{3-2\epsilon}2E_g}\frac{d^{3-2\epsilon}l}{(2\pi)^{3-2\epsilon}2l_0}\delta^{4-2\epsilon}(p_1+p_2-l-k)\nonumber\\
&=&\frac{(16\pi)^{-1+\epsilon}}{\Gamma(1-\epsilon)}\frac{(1-\tau)^{1-2\epsilon}}{\hat{s}^\epsilon}\int^\pi_0 d\theta\sin^{1-2\epsilon}\theta.
\label{eq:PS}
\end{eqnarray}

\par
To illustrate the cancellation of soft divergences and factorization of collinear divergences, the phase-space slicing method \cite{Baer:1989xj,Harris:2001sx} is used.  The soft region is defined by
\begin{eqnarray}
E_g<\frac{\sqrt{\hat s}}{2}x_\text{min},
\label{eq:soft}
\end{eqnarray}
and the collinear region is defined by
\begin{eqnarray}
1-|\cos\theta|<\delta, \label{eq:coll}
\label{eq:col}
\end{eqnarray}
where $E_g=\sqrt{\hat s}(1-\tau)/2$ is the gluon energy, and $x_\text{min}$ and $\delta$ are small arbitrary parameters.

\subsubsection{Soft region}
 Using the upperbound on final state gluon energy in Eq.~(\ref{eq:soft}), the contribution to the NLO cross section from soft gluon emission is
\begin{align}
\hat\sigma_\text{soft}&=\frac{1}{2\hat s}\frac{1}{4N^2_C}\int dPS^\epsilon_2|\mathcal{M}_{qq}|^2\Theta(\frac{\sqrt{\hat s}}{2}x_\text{min}-E_g)\nonumber\\
&=\frac{\sigma_0}{\hat s}\delta(1-\tau)\frac{\alpha_s}{2\pi}C_{\epsilon}\bigg{[}\begin{aligned}[t]
&2C_F\bigg{(}\frac{1}{\epsilon^2}-\frac{2\ln x_\text{min}}{\epsilon}+2\ln^2x_\text{min}-\frac{\pi^2}{6}\bigg{)}\\
&+C_D\bigg{(}\frac{1}{\epsilon}-2\ln x_\text{min}+2\bigg{)}\bigg{]}.
\end{aligned}
\end{align}
The single $\epsilon$ poles in the diquark color dependent corrections and the double $\epsilon$ poles cancel between the vertex corrections of Eq.~(\ref{eq:vet}) and soft gluon radiation.  Only the collinear divergences are left.

\subsubsection{Collinear region}
The collinear contribution to the cross section is then given by
\begin{eqnarray}
\hat\sigma_\text{col}&=&\frac{1}{2\hat s}\frac{1}{4N^2_C}\int dPS^\epsilon_2|\mathcal{M}_{qq}|^2\Theta(E_g-\frac{\sqrt{\hat s}}{2}x_\text{min})\Theta(\delta-1+|\cos\theta|)\nonumber\\
&=&\frac{\alpha_sC_F}{\pi}C_{\epsilon}\bigg{(}-\frac{1}{\epsilon}+\ln\frac{\delta}{2}\bigg{)}\int^{1-x_\text{min}}_0dx\,\sigma_\text{Born}(x\hat s)\frac{1+x^2-\epsilon(1-x)^2}{(1-x)^{1+2\epsilon}},
\end{eqnarray}
where $1-x$ is the energy fraction of the gluon with respect to an initial state quark and the lower bound $E_g>\frac{\sqrt{\hat s}}{2}x_\text{min}$ is imposed to avoid double counting with the soft gluon emission contribution.
\par
Performing the $x$-integration, the contribution to the real-gluon emission cross section from the collinear region is 
\begin{eqnarray}
\hat\sigma_\text{col}&=&\frac{\alpha_sC_F}{\pi}C_{\epsilon}\bigg{\{}\sigma_\text{Born}(\hat s)\bigg{[}\frac{2\ln x_\text{min}}{\epsilon}-2\ln\frac{\delta}{2}\ln x_\text{min}-2\ln^2x_\text{min}\bigg{]}\\
&&+\int^1_0dx\,\sigma_\text{Born}(x\hat s)\bigg{[}\bigg{(}-\frac{1}{\epsilon}+\ln\frac{\delta}{2}\bigg{)}\frac{1+x^2}{(1-x)_+}\nonumber\\
&&+(1-x)+2(1+x^2)\bigg{(}\frac{\ln(1-x)}{1-x}\bigg{)}_+\bigg{]}\bigg{\}},\nonumber
\end{eqnarray}
where the ``plus distribution'' is defined by
\begin{eqnarray}
\int^1_0dx g(x)\big{[}F(x)\big{]}_+=\int^{1-\beta}_0dx\,g(x)F(x)-g(1-\beta)\int^{1-\beta}_0dy\,F(y).
\end{eqnarray}

Adding the contributions calculated so far one obtains
\begin{multline}
\sigma_\text{Born}+\hat\sigma_\text{vertex}+\hat\sigma_\text{soft}+\hat\sigma_\text{col}\\
=\frac{\sigma_0}{\hat s}\bigg{[}\begin{aligned}[t]
&\delta(1-\tau)+\frac{\alpha_s}{2\pi}C_{\epsilon}\bigg{\{}2C_F\bigg{(}-\frac{3}{2\epsilon}-\frac{3}{2}\ln\frac{\hat s}{\mu^2_R}-1+\frac{\pi^2}{3}-2\ln\frac{\delta}{2}\ln x_\text{min}\bigg{)}\delta(1-\tau)\\
&\qquad+C_D\bigg{(}1-2\ln x_\text{min}-\frac{2\pi^2}{3}\bigg{)}\delta(1-\tau)\\
&\qquad+2C_F\bigg{[}\bigg{(}-\frac{1}{\epsilon}+\ln\frac{\delta}{2}\bigg{)}\frac{1+\tau^2}{(1-\tau)_+}+(1-\tau)+2(1+\tau^2)\bigg{(}\frac{\ln(1-\tau)}{1-\tau}\bigg{)}_+\bigg{]}\bigg{\}}\bigg{]}.
\end{aligned}
\end{multline}
The soft and UV divergences have been cancelled, hence the remaining divergences are collinear.  These collinear divergences can be absorbed into the definition of the pdfs. Using the \MSbar\ scheme, the universal counterterm for the collinear singularities is
\begin{eqnarray}
\hat\sigma^\text{CT}_\text{col}=\frac{\sigma_0}{\hat s}\frac{\alpha_s}{\pi}\bigg{(}\frac{4\pi\mu^2_R}{\mu^2_F}\bigg{)}^\epsilon\frac{1}{\Gamma(1-\epsilon)}P_{qq}(\tau),
\end{eqnarray}
where $\mu_F$ is the factorization scale and $P_{qq}(\tau)=C_F[(1+\tau^2)/(1-\tau)]_+$ is the DGLAP splitting function.
The sum of all the contributions to the NLO cross section so far calculated is
\begin{multline}
\sigma_\text{Born}+\hat\sigma_\text{vertex}+\hat\sigma_\text{soft}+\hat\sigma_\text{col}+\hat\sigma^\text{CT}_\text{col}\\
=\frac{\sigma_0}{\hat s}\bigg{\{}\begin{aligned}[t]
&\delta(1-\tau)+\frac{\alpha_s}{2\pi}\delta(1-\tau)\bigg{[}2C_F\bigg{(}\frac{\pi^2}{3}-1-2\ln\frac{\delta}{2}\ln x_\text{min}+\frac{3}{2}\ln\frac{\mu^2_R}{\mu^2_F}\bigg{)}\label{eq:NLOqq}\\
&\qquad+C_D\bigg{(}1-2\ln x_\text{min}-\frac{2\pi^2}{3}\bigg{)}\bigg{]}\\
&\qquad+\frac{\alpha_sC_F}{\pi}\bigg{[}\ln\bigg{(}\frac{\delta}{2}\frac{\hat s}{\mu^2_F}\bigg{)}\frac{1+\tau^2}{(1-\tau)_+}+(1-\tau)+2(1+\tau^2)\bigg{(}\frac{\ln(1-\tau)}{1-\tau}\bigg{)}_+\bigg{]}\bigg{\}}.
\end{aligned}
\end{multline}
The remaining dependence on the arbitrary parameters $x_\text{min}$ and $\delta$ will be removed by calculating the contribution from hard scattering.

\subsubsection{Hard scattering}
The remaining portion of the two-particle phase space, given by the constraints $E_g>\frac{\sqrt{\hat s}}{2}x_\text{min}$ and $1-|\cos\theta|>\delta$, corresponds to hard gluons in the final state.  The contribution to the cross section from hard scattering is 
\begin{eqnarray}
\hat\sigma_\text{Hard}&=&\frac{1}{2\hat s}\frac{1}{4N^2_C}\int dPS^\epsilon_2|\,\mathcal{M}_{qq}|^2\Theta(1-\delta-|\cos\theta|)\Theta(E_g-\frac{\sqrt{\hat{s}}}{2}x_\text{min})\\
&=&\frac{\alpha_s}{2\pi}\bigg{(}-2C_F\ln\frac{\delta}{2}-C_D\bigg{)}\int^{1-x_\text{min}}_0dx\,\sigma_\text{Born}(x\hat s)\frac{1+x^2-\epsilon(1-x)^2}{(1-x)^{1+2\epsilon}}\nonumber\\
&=&\frac{\sigma_0}{\hat s}\frac{\alpha_s}{2\pi}\bigg{(}2C_F\ln\frac{\delta}{2}+C_D\bigg{)}\bigg{\{}2\ln x_\text{min}\delta(1-\tau)-\frac{1+\tau^2}{(1-\tau)_+}\bigg{\}}.\nonumber
\end{eqnarray}
Adding all the cross sections together, the total NLO cross section with $qq$ initial state is

\begin{eqnarray}
\hat\sigma_{qq}&=&\frac{2\pi N_D{\lambda}^2(\mu^2_R)}{N^2_C\hat s}\bigg{[}\delta(1-\tau)+\frac{\alpha_s}{2\pi}\bigg{\{}2P_{qq}(\tau)\ln\frac{m^2_D}{\mu^2_F\tau}\\
&&+2C_F\bigg{[}2(1+\tau^2)\bigg{(}\frac{\ln(1-\tau)}{1-\tau}\bigg{)}_+ +\bigg{(}\frac{\pi^2}{3}-1-\frac{3}{2}\ln\frac{m^2_D}{\mu^2_R}\bigg{)}\delta(1-\tau)+1-\tau\bigg{]}\nonumber\\
&&-C_D\bigg{[}\frac{1+\tau^2}{(1-\tau)_+}+\bigg{(}\frac{2}{3}\pi^2-1\bigg{)}\delta(1-\tau)\bigg{]}\bigg\}\bigg{]}.\nonumber
\end{eqnarray}

\subsection{Gluon initiated process}
Next-to-leading order QCD corrections also arise from the gluon-initiated process
\begin{eqnarray}
g(p_1)+q(p_2)\rightarrow\bar{q}(k)+D(l),
\end{eqnarray}
shown in Fig.~\ref{fignloqg}.
The invariant amplitude for this process is
\begin{multline}
\mathcal{M}_{gq}=-2\sqrt{2}i g_s\mu^{\epsilon}_R\epsilon^A_{\mu}(p_1)\phi_{i}{u}^\trans(p_2)C^\dagger (\lambda_RP_R+\lambda_LP_L)\\
\times\bigg{[}t_A^\trans K^{i}\gamma^\mu\frac{\slashed p_1+\slashed p_2}{\hat s}-K^{i}{t_{A}}\frac{\slashed p_1-\slashed k}{2k\cdot p_1}\gamma^\mu-K^{j}T_A{}^i_j\frac{(2l-p_1)^\mu}{2l\cdot p_1}\bigg{]}u(k),
\end{multline}
where, again, the diquark color indices are left explicit while the quark color indices are suppressed.  Performing the spin sums the invariant amplitude squared is found to be
\begin{multline}
|\mathcal{M}_{gq}|^2=8\lambda^2 g^2_sN_D\mu^{2\epsilon}_R\bigg{\{}C_F\bigg{[}\frac{4}{1-\cos\theta}\bigg{(}\frac{1-\epsilon}{1-\tau}-2\tau\bigg{)}-(3+\cos\theta)(1-\tau)\bigg{]}\label{eq:gqamp}\\
+2C_D\bigg{[}1-\frac{4\tau}{(1+\tau)(1+\beta\cos\theta)}+\frac{8\tau^2}{(1+\tau)^2(1+\beta\cos\theta)^2}\bigg{]}\bigg{\}},
\end{multline}
where $\beta=(1-\tau)/(1+\tau)$, and $\theta$ is the angle between the initial state gluon and the final state antiquark.  The cross section for the gluon-initiated process is then
\begin{eqnarray}
\widetilde{\sigma}=\frac{1}{4(1-\epsilon)N_C(N^2_C-1)}\frac{1}{2\hat{s}}|\mathcal{M}_{gq}|^2dPS^\epsilon_2.
\end{eqnarray}

There are collinear divergences but no soft divergences in Eq.~(\ref{eq:gqamp}).  Following the same method as for gluon emission, we isolate the collinear divergences using the cutoff in Eq.~(\ref{eq:coll}).  The contribution to the gluon-initiated production cross section from the collinear piece is given by
\begin{eqnarray}
\widetilde{\sigma}_\text{col}&=&\frac{1}{4(1-\epsilon)N_C(N^2_C-1)}\frac{1}{2\hat{s}}\int dPS^\epsilon_2|\mathcal{M}_{gq}|^2\Theta(\delta-1+|\cos\theta|)\\
&=&\frac{\sigma_0}{\hat s}\frac{\alpha_s}{4\pi}C_{\epsilon}\bigg{\{}1+\bigg{[}(1-\tau)^2+\tau^2\bigg{]}\bigg{[}-\frac{1}{\epsilon}+\ln\frac{\delta}{2}+2\ln(1-\tau)-1\bigg{]}\bigg{\}}.\nonumber
\end{eqnarray}

Again, the collinear singularity can be absorbed into the definition of the pdfs.  Using the $\overline{\text{MS}}$ scheme, the universal collinear counterterm is
\begin{eqnarray}
\widetilde{\sigma}^\text{CT}_\text{col}=\frac{\sigma_0}{\hat s}\frac{\alpha_s}{2\pi}\bigg{(}\frac{4\pi\mu^2_R}{\mu^2_F}\bigg{)}^{\epsilon}\frac{1}{\Gamma(1-\epsilon)}\frac{1}{\epsilon}P_{qg}(\tau),
\end{eqnarray}
where $P_{gq}(\tau)=\frac{1}{2}[(1-\tau)^2+\tau^2]$ is the DGLAP splitting function.
Adding the collinear piece to the counterterm, we have
\begin{eqnarray}
\widetilde{\sigma}_\text{col}+\widetilde{\sigma}^\text{CT}_\text{col}=\frac{\sigma_0}{\hat s}\frac{\alpha_s}{4\pi}\bigg{\{}1+\bigg{[}(1-\tau)^2+\tau^2\bigg{]}\bigg{[}\ln\frac{\delta\hat s}{2\mu^2_F}+2\ln(1-\tau)-1\bigg{]}\bigg{\}}.
\end{eqnarray}

The hard scattering cross section does not contain any singularities and can be calculated in 4 dimensions.  The contribution to the cross section from hard gluon-quark scattering is then
\begin{align}
\widetilde{\sigma}_{\rm{Hard}}&=\frac{1}{4N_C(N^2_C-1)}\int{dPS_2}|\mathcal{M}_{gq}|^2\Theta(1-\delta-|\cos\theta|)\nonumber\\
&=\frac{\sigma_0}{\hat s}\frac{\alpha_s}{4\pi}\bigg{\{}\begin{aligned}[t]
&-\frac{3}{2}(1-\tau)^2-\ln\frac{\delta}{2}\bigg{[}(1-\tau)^2+\tau^2\bigg{]}\\
&+\frac{C_D}{C_F}\bigg{[}(1-\tau)(1+2\tau)+2\tau\ln\tau\bigg{]}\bigg{\}}.
\end{aligned}
\end{align}

Adding all the contributions together, we obtain the gluon-initiated cross section
\begin{align}
\hat{\sigma}_{gq}&=\widetilde{\sigma}_\text{col}+\widetilde{\sigma}^\text{CT}_\text{col}+\widetilde{\sigma}_{\rm{Hard}}\nonumber\\
&=\frac{N_D{\lambda}^2(\mu^2_R)\alpha_s}{2N^2_C\hat{s}}\bigg{\{}\begin{aligned}[t]
&2P_{gq}(\tau)\bigg{[}\ln\frac{m^2_D}{\mu^2_F\tau}+2\ln(1-\tau)\bigg{]}+\frac{(1-\tau)(7\tau-3)}{2}\\
&+\frac{C_D}{C_F}\bigg{[}(1-\tau)(1+2\tau)+2\tau\ln\tau\bigg{]}\bigg{\}}.
\end{aligned}
\end{align}

\bibliographystyle{h-physrev}
\bibliography{diquarkNLO}                      
\end{document}